\documentclass[preprint]{aastex}
\shorttitle{Star Formation in the Radio Galaxy NGC 4410A}
\shortauthors{Donahue, Smith \& Stocke}

\newcommand{\msun}{\rm{M}_\odot}

\newcommand{\halpha}{H$\alpha$}

\begin{document}
 
\title{Star Formation in the Radio Galaxy NGC 4410A}
 
\author{Megan Donahue\altaffilmark{1}}
\affil{Space Telescope Science Institute, 3700 San Martin Drive, Baltimore, MD
21218, donahue@stsci.edu}
\author{Beverly J. Smith\altaffilmark{1}}
\affil{East Tennesse State University, Dept. of Physics and
Astronomy, Box 70652, Johnson City, TN 37614, beverly@panda.etsu.edu}
\and
\author{John T. Stocke\altaffilmark{1}}
\affil{University of Colorado, CASA, CB 389, Boulder, CO 80309, 
stocke@casa.Colorado.edu}
\altaffiltext{1}{Visiting Astronomer, Kitt Peak National Observatory}

\begin{abstract}
The NGC~4410 group of galaxies provides us a rare opportunity to study 
a nearby (97 $h_{75}^{-1}$ Mpc) example of a radio galaxy (NGC~4410A) 
embedded in an extended X-ray source, with evidence for star formation 
that can be readily 
spatially distinguished from regions dominated by the AGN and shocks.
We present broadband and narrowband optical images along with
optical and IUE ultraviolet spectroscopy for the radio galaxy
NGC 4410A and its companion NGC 4410B.   Our H$\alpha$+[N~II] images reveal
six luminous H~II regions (L$_{H\alpha}$ $\sim$ 10$^{40}$ erg s$^{-1}$)
distributed in an arc near NGC 4410A.  
Partially completing 
the ring is a prominent stellar loop containing diffuse 
ionized gas.  This filamentary gas, in contrast to the H~II regions,
shows spectroscopic signatures of shock ionization.
The star formation in this system may have been triggered by a collision
or interaction between the two galaxies, perhaps by an expanding
density wave, as in classical models of ring galaxies.
Alternatively, the star formation may have been induced by
the impact of a radio jet on the interstellar matter.
Extended Ly$\alpha$ is detected in the ultraviolet IUE 
spectrum. The ultraviolet continuum,
which is presumably radiated by the nucleus of NGC~4410A, is
not extended. NGC~4410A appears to be interacting with its 
neighbors in the NGC~4410 group, and could be an example of a 
spiral galaxy transforming into an elliptical.

\end{abstract}
\keywords{galaxies:active, galaxies:interactions, 
galaxies:individual(NGC~4410A, NGC~4410B), X-rays, infrared radiation,
radio continuum, ultraviolet emission, HII regions, quasars:emission lines}

\section{INTRODUCTION \label{Intro}}

We do not understand the link between the radio sources in 
active galactic nuclei
and the emission line regions surrounding those sources. The
extended emission line regions around powerful radio galaxies
may be radiating light scattered from a central anisotropic source
(e.g. Tadhunter et al. 1993; Fabian 1989). Alternatively this gas
could be heated by shocks produced by the radio jet interacting with
the interstellar medium (ISM), or some combination of shock heating and
star formation (Rees 1989; Begelman \& Cioffi 1989; De Young 1989). In
the centers of cooling flow clusters, the radio galaxy morphology seems to
be intimately related to the structures seen radiating optical and
near-infrared emission lines typical of star-forming regions or
gentle shocks (Donahue et al. 2000; Koekemoer et al. 2000). In
the cooling flow clusters, the radio source seems to either excavate a 
cavity in the emission-line gas or the emission-line gas 
forms around the radio cavity. Similar structures are seen at high
resolution in the latest Chandra images of the central elliptical galaxies in
cluster cooling flows such as Hydra A (McNamara et al. 2000). Given the
variety of environments of these emission line regions, more 
than one process may be responsible for the emission-line 
nebulae in radio galaxies and central galaxies in clusters.

To investigate these issues, we made detailed
studies of a nearby radio galaxy in a group of galaxies that is  
close enough to resolve individual H~II regions, 
the peculiar low-luminosity radio galaxy NGC 4410A.
This system is quite nearby (97 $h_{75}^{-1}$ Mpc for $H_0 = 75h_{75}$ km s$^{-1}$ Mpc$^{-1}$),
and so provides a good target for detailed studies.
NGC 4410A, which is 
classified as an Sab? Pec galaxy on the basis of its
optical appearance (de Vaucouleurs et al. 1991),
has a prominent bulge surrounded by an extended ring or loop (Hummel,
Kotanyi, $\&$ van Gorkom 1986).  It forms a close
pair with the nearby 
S0? Pec galaxy NGC 4410B.  
These two galaxies are part of a sparse group containing
a dozen known members (Smith 2000). The velocity dispersion of those
members is $\sim220\pm70$ km s$^{-1}$, from the velocities
reported in Smith (2000), and the group has no obvious group 
emission (Tsch\"oke et al. 1999). Therefore this group is a poor, X-ray
faint group (e.g. Zabludoff \& Mulchaey 1998), a classification that is 
consistent with its lack of elliptical galaxies. 
The radio luminosity of NGC 4410A is near the faint end of the luminosity range 
for radio galaxies (Condon, Frayer, \& Broderick 1991), with a
4.8 GHz luminosity of 1.5 $\times$ 10$^{23} h_{75}^{-2}$
W Hz$^{-1}$.
The far-infrared 
(42.5 $-$ 122.5 $\mu$m) luminosity 
of the NGC 4410A+B pair is also moderate, 3.9 $\times$ 10$^9 h_{75}^{-2} \rm{L}\sun$ 
(Mazzarella,
Bothun, \& Boroson 1991, calculated as in Lonsdale
et al. 1985).  
The $L_{FIR}/L_{4.8~\rm{GHz}}$ 
ratio for this system is about 200 times less than expected for
a galaxy dominated by star formation, confirming
that NGC 4410A contains a radio-loud active nucleus
(Condon et al. 1991), in spite of its optical
classification as a disk galaxy.

NGC 4410A has a very peculiar
radio morphology: a shorter (3$'$ $\sim 80 h_{75}^{-1}$ kpc) very distorted radio lobe
to the southeast
(Hummel et al. 1986; Batuski et al. 1992)
and a fainter longer structure (7$'$ $\sim 200 h_{75}^{-1}$ kpc) to the northwest 
(Smith 2000).   
This strange radio morphology may have been caused by an interaction of the radio
lobe with interstellar matter that has been disturbed by a gravitational
encounter 
between NGC 4410A and NGC 4410B
(Smith 2000).
NGC 4410A+B contains 
abundant interstellar matter ($M_{HI} \sim 10^9 h_{75}^{-2} \msun$ and
$M_{H_2} \sim 4 \times 10^9h_{75}^{-2} \msun$ (Smith 2000; assuming the standard Galactic
$I_{CO}/N_{H_2}$ conversion factor.) 
About a third of the HI
lies in a tail-like structure extending 1\farcm7 ($50 h_{75}^{-1}$ kpc) to the 
southwest, coincident with a faint optical tail (Smith 2000).
This tail overlaps with the southeastern radio lobe, suggesting an interaction
between the radio jet and the HI gas.

In addition to an active nucleus, NGC 4410A  has on-going 
star formation.
In this paper, we present evidence for luminous H~II regions
in NGC 4410A, in the form of new optical images
and optical and ultraviolet spectroscopy.

\section{OBSERVATIONS}

\subsection{Optical Imaging}

We observed NGC~4410 on 1989 March 7 using the 512$\times$512 Tek \#1 CCD 
direct camera mounted on the Kitt Peak National Observatory 2.1m 
telescope.  This system provided a pixel size of 0\farcs34, giving
a field of view of 2\farcm9. 
The seeing was 
excellent ($\sim$0\farcs9) and the sky was photometric.  
We obtained two 15 
minute exposures through an on-band \halpha\ filter centered on the 
redshift of the cluster ($\lambda_o = 6737 {\rm \AA}$, $\Delta \lambda = 75 
{\rm \AA}$), and a single 15 minute  exposure through an off-band filter
centered 100${\rm \AA}$ blueward ($\lambda_o = 6606 {\rm \AA}$, $\Delta
\lambda = 75 {\rm \AA}$).  The [N~II] $\lambda$6548,6584${\rm \AA}$ lines
are included in the bandpass of the on-band filter.
These images were calibrated by observations of 
Hiltner 600 made the same night with the same filters and similar air masses.
In order to obtain a pure emission-line image of the galaxy, after
sky subtraction we 
used the three brightest stars in the field to derive
the ratio of transmission through the on-band and off-band filters. 
The off-band image was scaled according to this ratio and 
subtracted from the on-band image.

A larger field of view R-band image of the inner portion of the NGC 4410 group
was obtained using the 
Southeastern 
Association for Research in Astronomy (SARA)
0.9m telescope on Kitt Peak.
These observations were made on 1999 April 18 $-$ 19, using a 2048 $\times$ 2048
Axiom/Apogee CCD.  Binning 2 $\times$ 2, 
the pixel size is 0\farcs52 
with this system
and
the field of view is 8\farcm9. The seeing was $\sim2"$.
The total integration time on the galaxy was 40 minutes.
Sky flats, darks, and biases were also obtained, and the data
were reduced in a standard way.  

In this paper, we also discuss an archival Hubble Space Telescope (HST)
Wide Field Planetary Camera 2 (WFPC2) image of NGC 4410A\footnote{Based
on observations made with
the NASA/ESA Hubble Space Telescope, obtained from the data archive
at the Space Telescope Science Institute.  STScI is operated by
the Association of Universities for Research in Astronomy, Inc. under
NASA contract NAS 5-26555.},
obtained on 1995 April 20.
These data consist of a single 500 second exposure taken with
the broadband red F606W HST image ($\lambda$$_0$ = 5934${\rm \AA}$;
$\Delta$$\lambda$ = 1498${\rm \AA}$).
The nucleus of NGC 4410A was centered in the high resolution Planetary
Camera (PC) chip of the WFPC2, which provides 800 $\times$ 800
0\farcs0455 pixels.
These data were originally obtained as part of the Malkan, Gorjian, \& Tam (1998)
HST survey of active galaxies, and have previously been presented
by Tsch\"oke et al. (1999) and Smith (2000).

\subsection{Optical Spectroscopy}

Optical longslit spectra of NGC 4410 were obtained on 
1991 December 19 $-$ 20 using the
Double Spectrograph mounted at the Cassegrain focus of the Palomar
5m Hale telescope. 
In the red
channel, a 1200 line mm$^{-1}$ grating blazed near 7000${\rm \AA}$ was used.  
This provided
a dispersion of 0.817${\rm \AA}$ pix$^{-1}$, a resolution of $\sim$ 
2.5${\rm \AA}$, and 
total wavelength coverage of 6340 $-$ 7000${\rm \AA}$.  
In the blue channel, we used a 300 line
mm$^{-1}$ grating, which gave a dispersion of 2.153${\rm \AA}$ pix$^{-1}$, 
a resolution of $\sim$6${\rm \AA}$, and a wavelength range of 3747 $-$ 
5250${\rm \AA}$.

\begin{figure}
\plotone{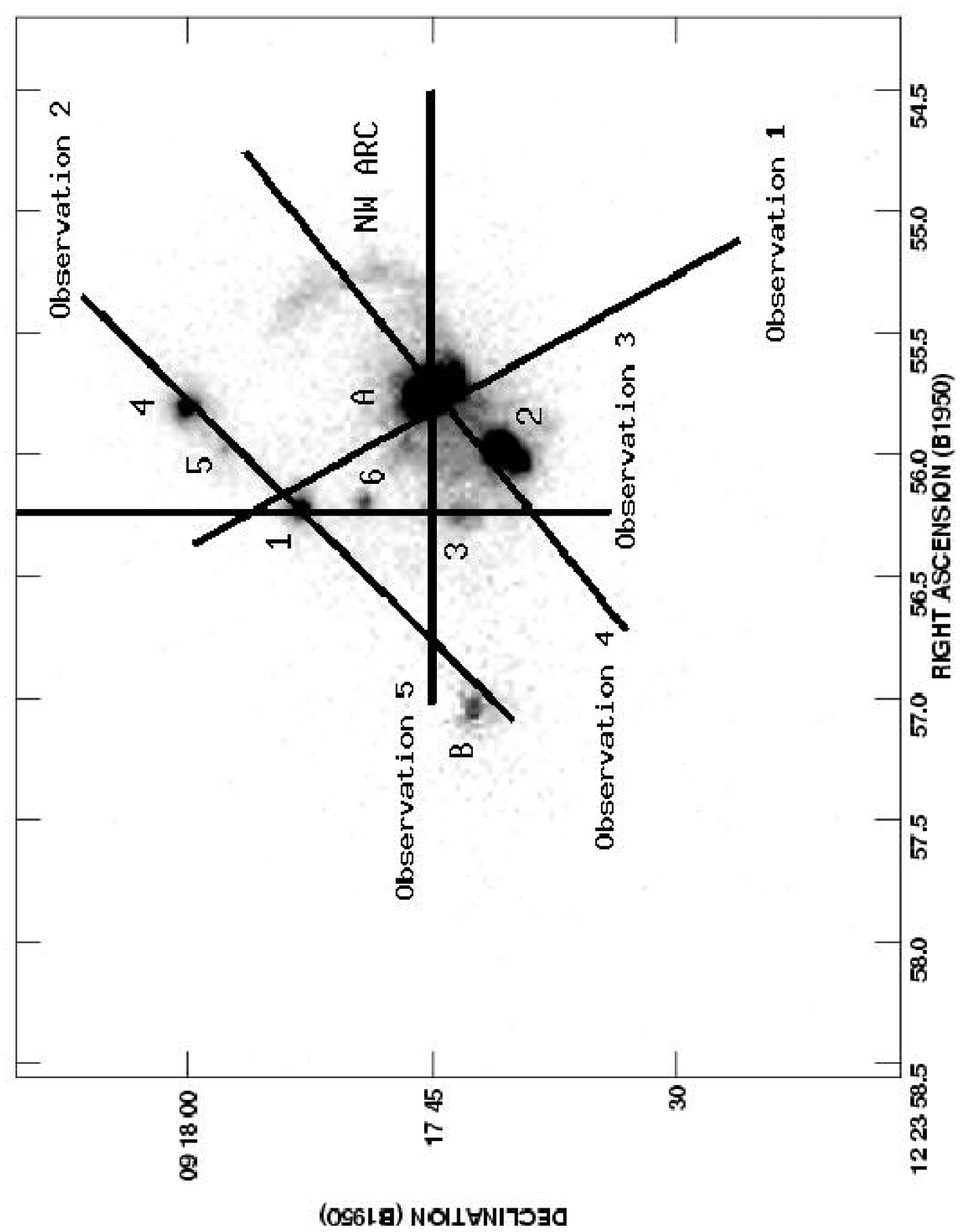}
 \caption{
The locations of the slit positions on NGC 4410 for the optical
spectroscopy, superposed on the H$\alpha$+[N~II] image. The actual
slit widths were 2" wide; the accuracy of the slit locations is
about 0.5-1.0", based on telscope slew offsets from a nearby star. 
Features like Knot \#2 and nucleus B were probably not well-centered
in the slit.  
The H~II regions tabulated in Table~\ref{Table1} are identified in this image.
\label{fig1}}
\end{figure}

We made observations 
at five slit positions in the NGC 4410 system, 
aligned across the nucleus of NGC 4410A and various knots and
features observed in the narrow-band image.  Slit positions are indicated
in Table~\ref{Table2} and are marked in Figure~\ref{fig1}.
Individual exposures were 15 minutes in duration.  
We used a 2$''$ slit for all observations. The seeing was 2$''$. 
All observations were conducted at less than 1.1 airmasses.

\subsection{Ultraviolet Spectroscopy}

Using the International Ultraviolet Explorer ({\em}IUE),
on 1990 January 30 and February 1
we obtained low-dispersion ultraviolet spectra of NGC~4410 
in a 10$''$ $\times$ 20$''$
slit 
centered on NGC 4410A.  The position angle of the slit 
was 
178$^{\circ}$ east of north.
These
observations covered a wavelength range of 1150 $-$ 1975${\rm \AA}$ 
with the Short Wavelength
Prime (SWP) Camera
and 1910 $-$ 3300${\rm \AA}$ with the Long Wavelength Prime (LWP) Camera.
The SWP 
exposure was 380 minutes in duration, and the LWP
exposure lasted 377 minutes.

\section{RESULTS}

\subsection{The Narrowband Images}

\begin{figure}
\plotone{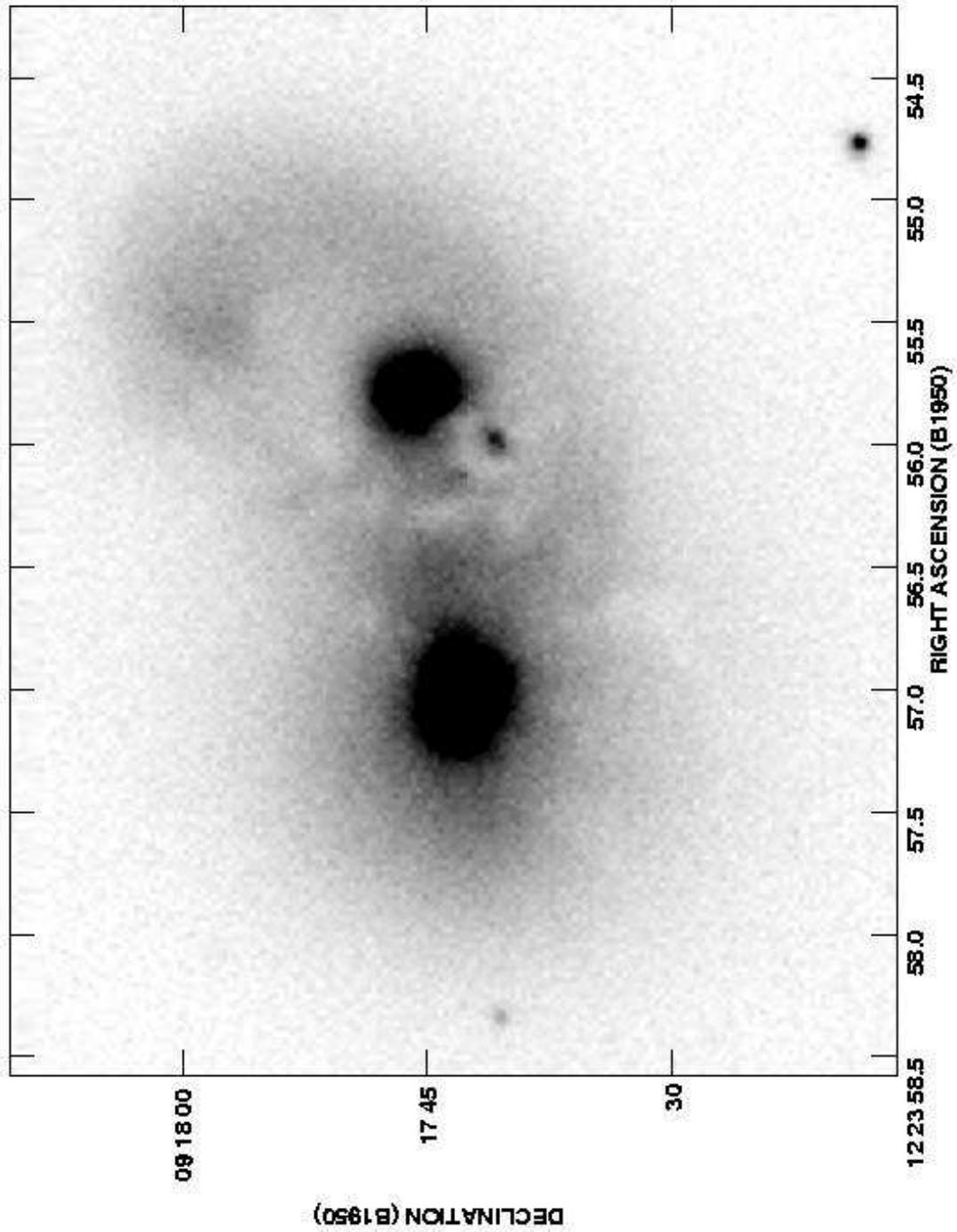}
 
\caption{
The inner 65$''$ $\times$ 54$''$ of the narrowband 6606{\rm $\AA$}
continuum image of NGC 4410A (to the west) and NGC 4410B.
Note the ring-like structure to the northwest of NGC 4410A and the
bright knot 5\farcs4 southeast of the nucleus of NGC 4410A.
\label{fig2}}
\end{figure}

\begin{figure}
\plotone{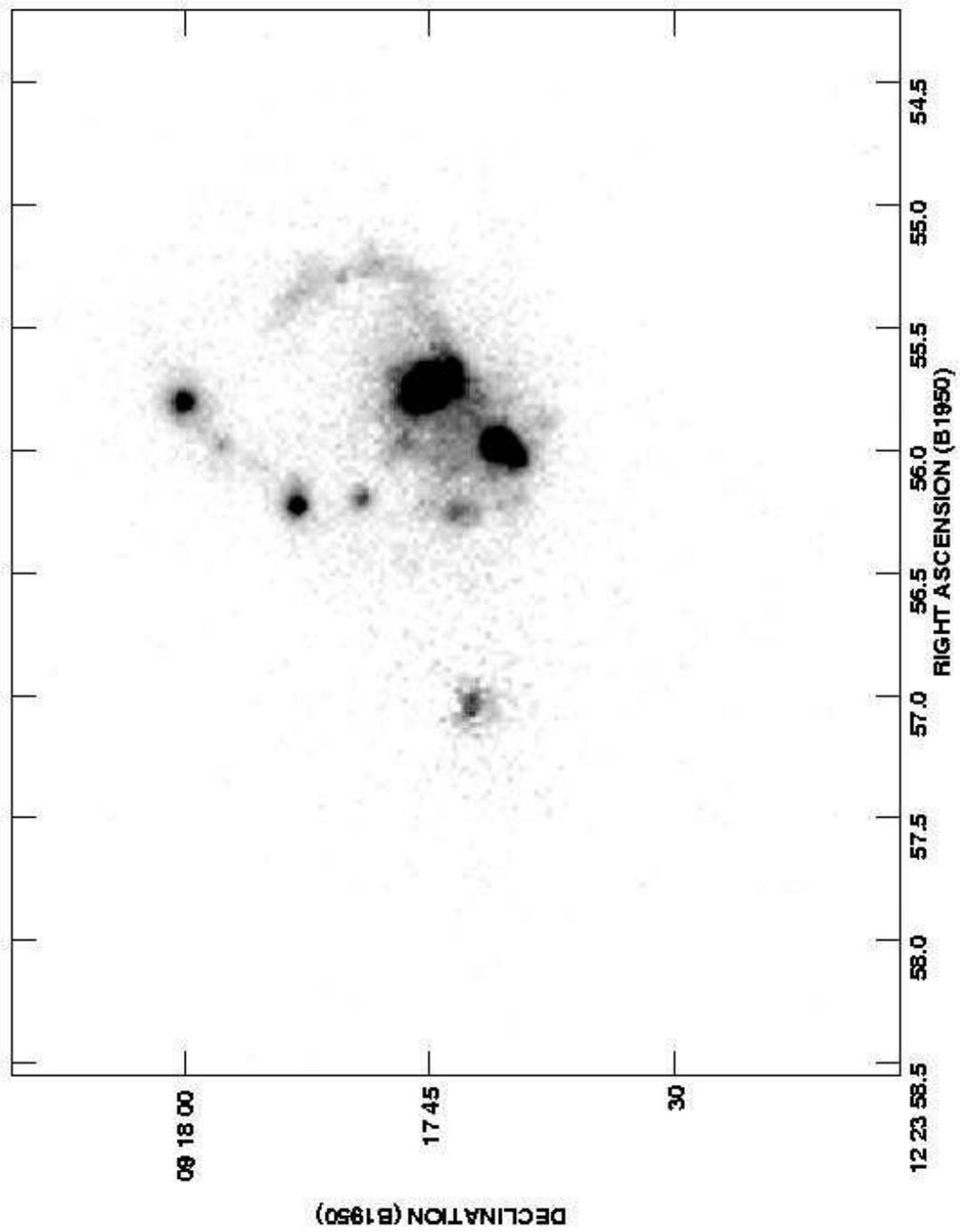}
 \caption{
The inner 65$''$ $\times$ 54$''$ of the continuum-subtracted
H$\alpha$+[N~II]
image of NGC 4410A and NGC 4410B.
Both galactic nuclei are visible in this image, as well as the
luminous knot southeast of the NGC 4410A nucleus,
a string of five H~II regions along the eastern
portion of the ring, and a filamentary structure along the western
section of the ring.
\label{fig3}}
\end{figure}

The $6606 {\rm \AA}$ continuum image of NGC 4410A and NGC 4410B
is displayed in Figure~\ref{fig2},
while the continuum-subtracted 
H$\alpha$+[N~II] map is presented in Figure~\ref{fig3}.
The ring-like structure of NGC 4410A is evident in both
the 
continuum and the line map.
In the continuum map, the western portion of the ring is more prominent,
with a bright concentration north of the NGC 4410A nucleus.
In the H$\alpha$+[NII] image, a series of unresolved luminous knots lie
along the eastern portion of the ring, while an arc of lower surface
brightness filamentary
emission is visible along the western section of the ring.
Both nuclei are detected in H$\alpha$+[N~II],
and another
extremely bright source is visible 5\farcs4 southeast of the 
nucleus.
This source 
is elongated northwest to southeast by 
$\sim$2\farcs5 (1.2 kpc), and 
is also visible in the continuum image (Figure~\ref{fig2}).
The nuclei of both galaxies
are resolved in the H$\alpha$+[N~II] map, with the center of NGC 4410A
being elongated $\sim$10$''$ southwest$-$northeast.
In the continuum image, prominent dust features are visible
between the two galaxies, near the bright knot southeast
of the NGC~4410A nucleus, and south of NGC 4410B.

The total H$\alpha$ + [N~II] flux from this system
is $3.7\pm1.2 \times 10^{-13}$ erg s$^{-1}$ cm$^{-2}$.
In Table\ref{Table1}, we provide positions and H$\alpha$+[N~II] fluxes for the
centers of the two galaxies and the six brightest knots,
along with a total flux for the northwestern arc. These sources
are labeled on the H$\alpha$+[N~II] map in Figure~\ref{fig1}. 
The center core of NGC 4410A is six times brighter in H$\alpha$+[N~II]
than the center of NGC 4410B, but only 1.8 times more luminous than
the most luminous knot, Knot $\#$2. 
The filamentary arc in the northwest provides $\sim$10$\%$
of the total H$\alpha$+[N~II] flux of the system.

Hummel et al. (1986) noted the presence of two optical
concentrations in the southeastern tail of NGC 4410A+B.
The more northern of these, at 12$^{\rm h}$ 23$^{\rm m}$ 58.8$^{\rm s}$
9$^{\circ}$ 16$'$ 14$''$ (1950) is in the field
of view of our narrowband images.
This source was unresolved in the continuum image
and was undetected in the 
H$\alpha$+[N~II]
map.
The 3$\sigma$ upper limit to the H$\alpha$+[N~II] flux from this
source 
is 2.9 $\times$ 10$^{-16}$ erg s$^{-1}$ cm$^{-2}$.

\subsection{The HST Image}

\begin{figure}
\plotone{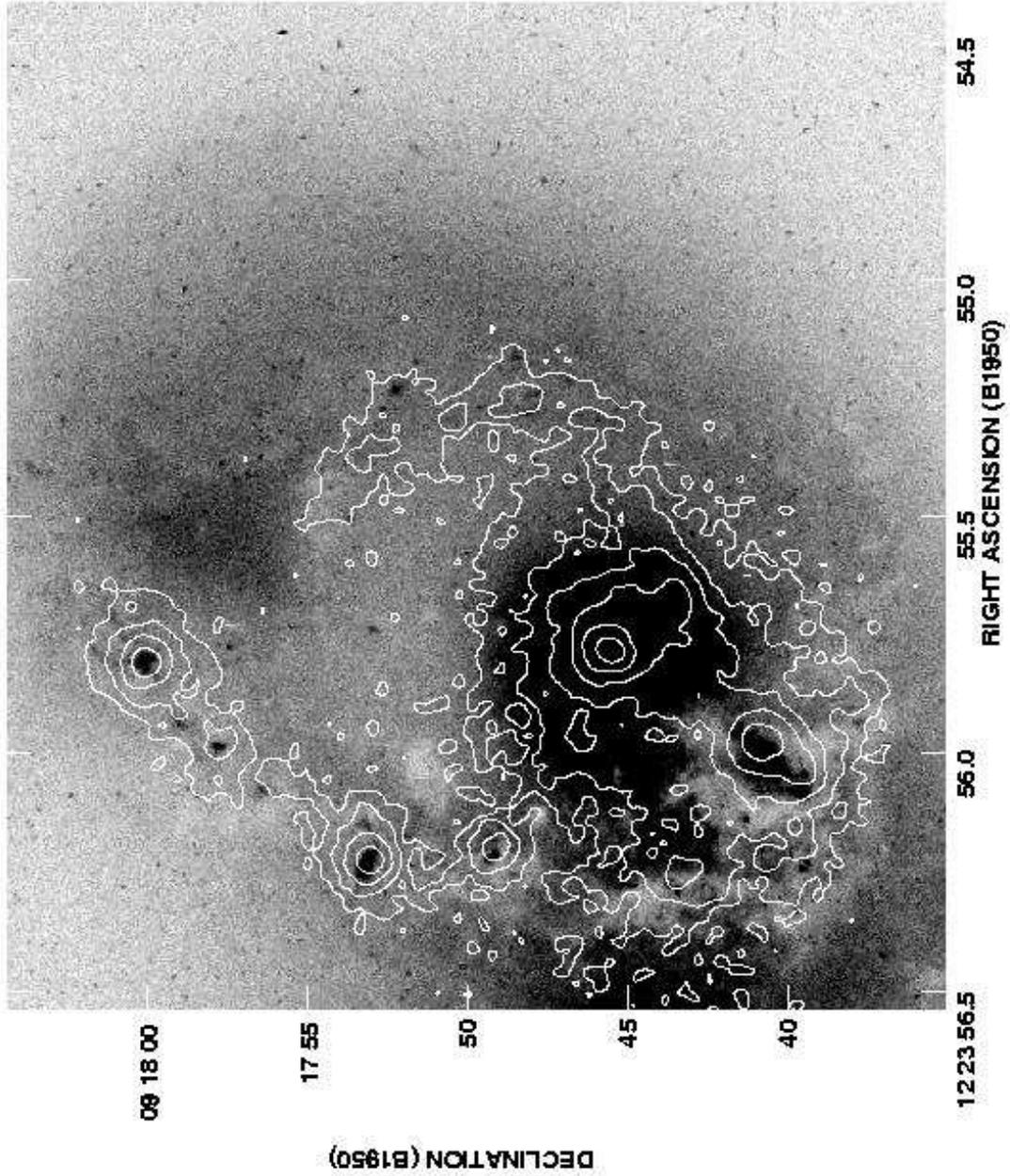}
 \caption{
The H$\alpha$+[N~II] image of NGC 4410 superposed 
on the HST broadband image.  The first contour is 4.1 
$\times$ 10$^{-16}$
erg s$^{-1}$ cm$^{-2}$.  The contours are logarithmic, increasing
by $\Delta$log F = 0.3.
\label{fig4}}
\end{figure}

In Figure~\ref{fig4}, the 
ground-based H$\alpha$+[N~II] map is superposed on
the archival HST image.  We report F606W photometry in
Vega magnitudes ($m_R$) for some of the 
features inside 0\farcs96 apertures. Since good cosmic ray rejection was
not possible and the features -- most of which are spatially 
resolved by HST --  are embedded in a complex sky background, the
photometry is not precise ($\pm 0.2$ magnitudes). 
A good coincidence is found
between the knots seen in the narrowband image
and clumps in the HST map.  In the HST image,
these knots are well-resolved.
Knot $\#$4 is extended 0\farcs36 $\times$ 0\farcs18 (170 $\times$ 85 pc)
east-west, with at least two subknots with $m_r\sim20.5$ each).  
Knot $\#$1 is elongated 
0\farcs59 $\times$ 0\farcs32 (278 $\times$ 150 pc)
northeast to southwest, and has 3 components, with an aperture
magnitude of $\sim 20.1$.
Knot $\#$3 has a larger angular size, 
$\sim$1\farcs5 $\times$ 1\farcs5 (700 pc
$\times$ 700 pc), but a lower surface brightness ($m_R\sim 20$ mag arcsec$^{-2}$,
 $m_R\sim 21$ in a 0.5 arcsec aperture).
  The most luminous knot external to the nucleus, 
Knot $\#$2, is very extended southeast to northwest,
with a size of 2\farcs67 $\times$ 1\farcs14
(1.26 $\times$ 0.54 kpc).  This knot has three prominent optical peaks
and at least six other subclumps.
The two brightest subknots ($m_R \sim 18.8$ magnitudes each) in the northeast are separated by 0\farcs69 (330 pc),
while the third brightest subknot ($m_R \sim 20.7$ magnitudes) 
lies 1\farcs14 (540 pc) to the southeast.
These subknots have sizes $\sim$ 0\farcs2 (90 pc).

The dust lanes seen in the groundbased red continuum image
are also visible in the HST image
and are well-resolved, with widths of $\sim$0\farcs4 $-$ 1\farcs4 (190 $-$ 660
pc).
These dust features have a patchy filamentary structure, 
and are visible on both sides of the loop, but are
more prominent in the eastern side and surrounding
Knot $\#$2. A particularly strong dust lane feature lies just to the southwest
of the nucleus, with a position angle of $\sim123$ degrees E of N. The
dust lane may obscure the actual nucleus since the peak of the optical
emission lies next to the dust lane feature.

\subsection{The Ground-Based Broadband Optical Images}

\begin{figure}
\plotone{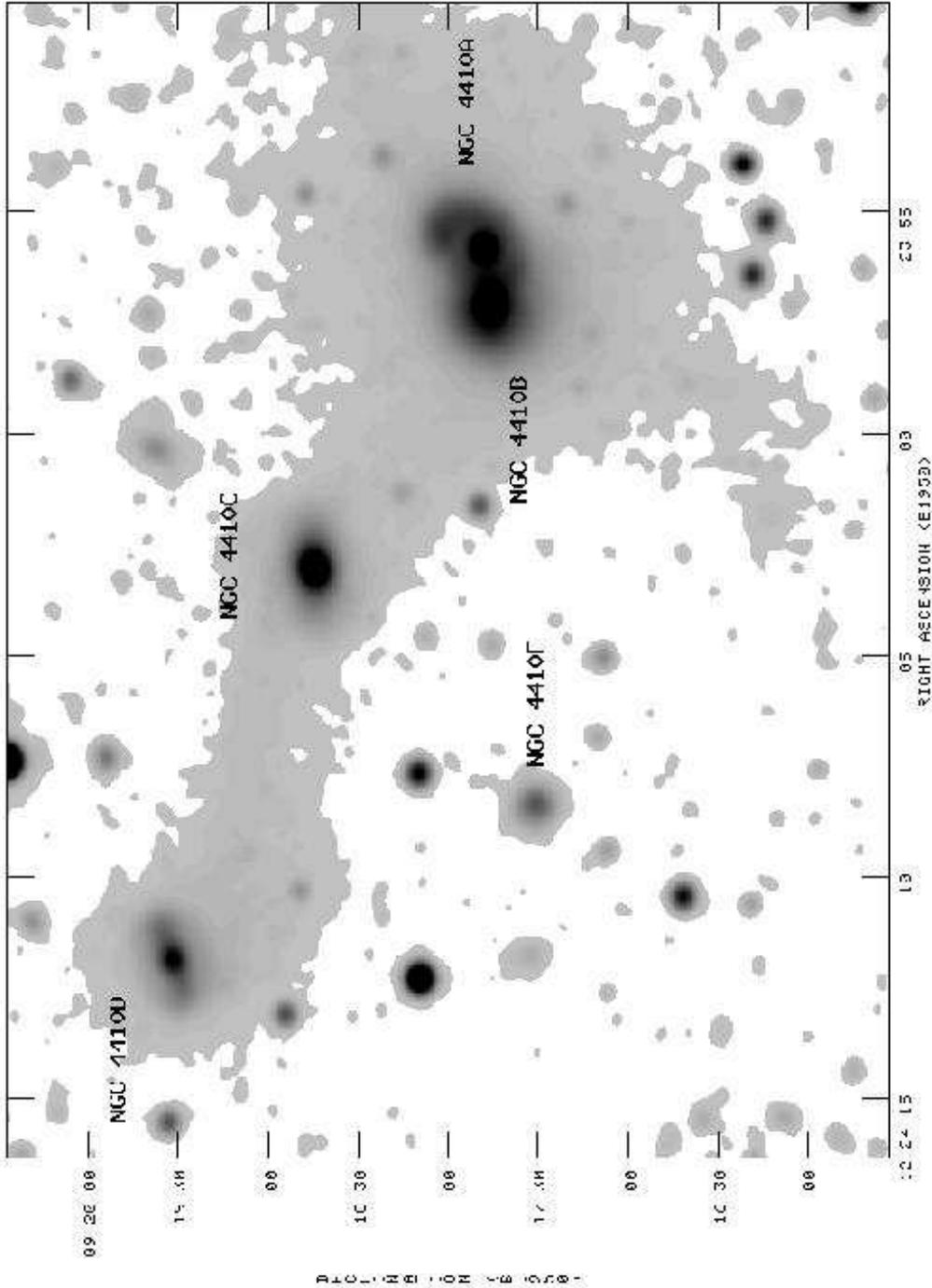}
 \caption{
The SARA R band image of NGC 4410.  
The five galaxies in the inner part of the NGC 4410 group are labeled.
This image has been smoothed with a 5$''$ Gaussian to emphasize the faint
features.
NGC 4410A, B, and C are classified as Sab? Pec, S0? Pec, and S0?
respectively, by de Vaucouleurs et al. (1991).  NGC 4410D is classified
as SBa(s) by the NASA Extragalactic Database (NED), and
NGC 4410F is classified as S by Smith (2000). 
\label{fig5}}
\end{figure}

In Figure~\ref{fig5}, the larger field of view R band image from the SARA
telescope is displayed. 
In this image, five of the galaxies in the inner part
of the group are visible: NGC 4410A, B, C, D, and F.
The long tail to the southeast of NGC 4410A+B previously noted 
by Hummel et al. (1986) is evident in this
image, as is the broad extension to the northwest of
this pair. 
Roughly calibrating this image using the galaxy magnitudes
from de Vaucouleurs et al. (1991) and
assuming B $-$ R $\sim$ 1.5, we estimate that the
two optical knots in the southeastern tail identified
by Hummel et al. (1986) have R magnitudes $\sim$ 19.

\subsection{The Optical Spectroscopy \label{opt_spec}}

From the long-slit spectra, we extracted $\sim$2$''$ $\times$ 4$''$
aperture measurements for the two nuclei, four of the H~II regions,
the northwestern arc, and the extended ionized gas east, west,
and southwest of the NGC 4410A nucleus.
These spectra are displayed in Figure~\ref{fig6}.
From these data, we extracted line ratios,
as well as the central velocity and line width (Table~\ref{Table3}).
The Galactic extinction in the direction of NGC4410 is very small,
with $A_B \sim 0.0$ (Burstein \& Heiles 1984), so
no intrinsic correction for Galactic absorption was applied to the spectra.
The uncertainty in the Galactic absorption introduces up to 
0.1-0.05 magnitudes of systemic 
uncertainty to relative line strengths, particularly
for [O~II]/[O~III] and H$\alpha$/H$\beta$.
Because the seeing was comparable to the slit width, the absolute
photometry is uncertain, so in Table~\ref{Table3} we only provide relative
line fluxes and statistical uncertainties. 
Knot \#2 does not have a blue spectrum because it was
unresolved from the nucleus in our observations, owing to the poorer
optics on the blue side of the Double Spectrograph. 

Since the observations were executed at low airmass ($<1.1$), the 
corrections for differential refraction (Filippenko 1982) would be 
$<10\%$ for the [OII]/H$\beta$ ratios, and 
$<2-3\%$ for the H$\alpha$/H$\beta$ ratios for the position angles
used. 
One of the observations, Observation 1 with a position angle of 25 degrees,
was very close to the parallactic angle of 33 degrees. The differences
between the ratios for Knot \#1 at this position angle exceed
that provided by differential refraction. 

When possible, the H$\alpha$/H$\beta$ ratio was used to estimate the extinction
to each region (Table~\ref{Table4}), 
assuming an intrinsic H$\alpha$/H$\beta$ ratio of
2.85 (case B) for the H~II regions and 3.1 (Ferland \& Netzer 1983)
for the NGC 4410A nucleus, along with additional, probably 
conservative, systematic uncertainties of 
12\% for the line ratios. A dust screen with the 
extinction curve of Seaton (1979) was assumed. 
In most cases, the implied extinction is moderate ($A_{\rm V} \sim 1-1.5$).
If stellar absorption at H$\beta$ relative to
that at H$\alpha$ is significant, the actual extinction from dust 
is somewhat lower. Stellar absorption can remove a few Angstroms
of equivalent width from the H$\beta$ emission line. The typical
equivalent width of H$\beta$ emission was (-)3-8\AA, so moderate
stellar absorption could cause an overestimate by 0.4-0.8 magnitudes 
of the absorption at H$\beta$.
We note that if the ionized gas and dust are well-mixed, 
the amount of dust present is higher than what is inferred from
extinction derived from assuming a screen. 

Comparison of the observed line ratios
(Table~\ref{Table3})
with standard diagonostic line-ratio diagrams used to distinguish
ionization mechanisms (e.g., Baldwin, Phillips, \& Terlevich
1985; Veilleux \& Osterbrock 1987) confirms that
the bright knots of ionized gas 
are H~II regions.
In contrast, the nuclei of the two galaxies fall in the
Low Ionization Emission Line Region (LINER) regime, consistent
with the classification of these galaxies' nuclear spectra by
Mazzarella \& Boroson (1993).
The
[N~II]6584/H$\alpha$ ratio is high in both nuclei,
and the 
[S~II]6717/H$\alpha$ and [O~I]6300/H$\alpha$ ratios 
are enhanced
in the NGC 4410A nucleus. The fainter NGC 4410B nucleus is undetected
in 
[S~II]6717 and [O~I]6300.
LINER-like spectra are common in radio galaxies (e.g., Baum 
\& Heckman
1989; Baum, Heckman, \& van Breugel 1990).
The northwestern arc
also has enhanced [N~II]6584/H$\alpha$, 
[S~II]6717/H$\alpha$, and [O~I]6300/H$\alpha$ ratios,
indicating that energy input to the gas in this region 
may be dominated by shocks or other
heating mechanisms  as opposed to 
the radiation from hot stars.
The extended emission to the southwest, east, and west of the 
NGC 4410A nucleus has ratios in between those of H~II regions
and LINER spectra, suggesting that light from both the nucleus
and nearby H~II regions may have contributed to the extracted spectra.

Additional evidence that we are distinguishing regions dominated by 
AGN activity from H~II regions
in NGC4410 comes from the line widths and the lack of strong 
continuum in the H~II sources.
The nucleus of NGC 4410A has relatively broad lines (FWHM $\sim$ 600
km s$^{-1}$).  The NGC 4410B nucleus and the filamentary
features also appear to have somewhat broad lines (FWHM $\sim$ 400
km~s$^{-1}$) compared to those of the knots
(FWHM $\sim$ 200 $-$ 300 km~s$^{-1}$), which are only marginally
resolved.  The knot sources have very little continuum, consistent
with their identification as H~II regions.

The velocity structure of the ionized gas in this
system (Figure~\ref{fig7}) is intriguing.  The two nuclei, the northwestern
filament, and the extended emission to the east of the NGC 4410A nucleus
are considerably redshifted (7440 km~s$^{-1}$ $-$ 7500 km~s$^{-1}$)
from the H~II regions and the extended emission to the
southwest and east of the NGC 4410A nucleus, centered at 
7130 $-$ 7340 km~s$^{-1}$). It is possible that we are viewing line
emission from two interacting systems with a velocity separation of
$\sim 200$ km s$^{-1}$, or that the H~II regions were originally 
associated with one of the nuclei.

\subsection{The Ultraviolet Spectra}

The IUE data were extracted and calibrated by the IUE
New Spectral Imaging Processing System 
(NEWSIPS)
software in two ways: the standard extraction method
which is optimized for a point source, 
and a re-extraction and reprocessing using a method
optimized for an extended source.
The two SWP spectra are
plotted in Figure~\ref{fig8}.

The 1216${\rm \AA}$ Ly$\alpha$ feature is strongly
detected in this source,
with a total line flux of 
1.0 $\pm$ 0.2 $\times$ 10$^{-13}$ 
erg s$^{-1}$ cm$^{-2}$. A possible  
1548+1551${\rm \AA}$ 
C~IV feature may be present at the level of 
6 $\pm$ 2 $\times$ 10$^{-14}$ 
erg s$^{-1}$ cm$^{-2}$, but this feature coincides with the
presence of some cosmic rays and the wavelength centroid of
the feature places it significantly bluer (by about 400 km sec$^{-1}$)
of the centroid of the Ly$\alpha$ feature, and lies about 4.5$''$ south
of the position of the continuum source. If real, it may be associated
with Knot $\#$2, but we are doubtful.

The mean continuum between 1350-1450 \AA~observed frame 
in the SWP spectrum is 
$1.1\pm0.2 \times 10^{-15}$ 
erg s$^{-1}$ cm$^{-2}$
{\rm \AA}$^{-1}$.
The continuum flux from the SWP spectrum obtained with the extended source
algorithm is statistically 
equal to that found with the point source algorithm 
centered on the 
continuum peak ($1.2\pm0.2 \times 10^{-15}$ erg s$^{-1}$ cm$^{-2}$ \AA$^{-1}$),
 indicating that the continuum
emission is unresolved by IUE 
(FWHM $<$ 3\farcs5).
The LWP spectrum (not shown) contains only faint, flat continuum emission, at
a mean level of 4 $\pm$ 2 $\times$ 10$^{-16}$ 
erg s$^{-1}$ cm$^{-2}$
{\rm \AA}$^{-1}$ at 3000{\rm \AA}.

Ly$\alpha$ is clearly visible in both extracted spectra,
and the flux obtained with the extended source method
is higher than that derived assuming a point source. 
In Figure~\ref{fig8}, the amplitude of the Ly$\alpha$ emission 
line from the extended source extraction ($\sim 1.2 \times 10^{-14}$  
erg s$^{-1}$ cm$^{-2}$
{\rm \AA}$^{-1}$) is higher than that obtained 
from the point-source extraction 
($\sim 0.5 \times 10^{-14}$
erg s$^{-1}$ cm$^{-2}$
{\rm \AA}$^{-1}$),
indicating that Ly$\alpha$
is extended.  
The extent of the Ly$\alpha$ emission is 
$>20''$, since the emission seems to fill the IUE aperture.

The velocity of the Ly$\alpha$ line is 7050 $\pm$ 50 km s$^{-1}$.
This velocity is inconsistent
with the optical velocity of the NGC 4410A nucleus, but similar
within the uncertainties with the optical velocities of the H~II regions
(Table~\ref{Table3}), indicating that much of the Ly$\alpha$ emission
may originate in  
the H~II regions rather than the nucleus, as expected from
the extended spatial distribution of the Ly$\alpha$ emission.

The total observed H$\alpha$ luminosity from the NGC 4410A+B system
is 2.2 $\times$ 10$^{41}$ erg~s$^{-1}$, after
correction for 
[N~II] using the spectroscopic results in Table~\ref{Table3} and assuming the
diffuse emission has line ratios similar to the northwest arc.
This gives an overall Ly$\alpha$/H$\alpha$ flux ratio for this system
of $\sim0.50$. The expected intrinsic 
Ly$\alpha$/H$\beta \sim 40$ from photoionization
calculations, but the observed ratio is affected by geometry and
resonant scattering effects.   
Assuming Case B recombination line ratios for H$\alpha$/H$\beta
\sim 2.8$ (Osterbrock 1989) and the extinction
curve of Seaton (1979), the observed Ly$\alpha$/H$\alpha$ ratio implies
an average extinction A$_{\rm V} \sim 1.7$ towards the system, somewhat  
higher than but less reliable (Giavalisco, Koratkar \& Calzetti 1996) 
than the extinctions estimated from the 
H$\alpha$/H$\beta$ ratios (Table~\ref{Table4}).

\section{DISCUSSION}

\subsection{Star Formation in NGC 4410}

Both star formation and non-thermal nuclear activity are
occuring in NGC 4410A, as well as possible shock-ionization of 
interstellar matter. Here we will discuss the evidence for all
of these phenomena, but in particular for 
on-going star formation in NGC 4410. We compare the star formation
evidence in this peculiar, perhaps intermediate morphological
type, galaxy with star formation signatures in other classes
of galaxies.

Our H$\alpha$+[N~II] map reveals the presence of both
compact knots and diffuse extended emission, as well
as extended emission associated with the two nuclei.
Our optical spectroscopy indicates that the knots are H~II regions, 
the filamentary structure to the northeast of the nucleus
may be shock-ionized, and the nuclei of NGC~4410A and NGC~4410B 
have LINER spectra. As noted in the Introduction, NGC 4410A 
has a peculiar double-lobed radio structure
(Smith 2000),
and a low
far-infrared to 4.8 GHz radio luminosity ratio, 
therefore NGC 4410A contains both a radio-loud active nucleus
and on-going star formation.

The H~II region complexes in NGC 4410 are very luminous.
Knot $\#$2 has L$_{H\alpha}$ = 3.4 $\times$ 10$^{40}$ erg s$^{-1}$
(corrected for [N~II]6584 using the [N~II]6584/H$\alpha$ ratio 
in Table~\ref{Table3}, but uncorrected for extinction).
This is twice as luminous as the `supergiant' H~II
region 30 Doradus in the Large Magellanic
Cloud (Kennicutt 1984).
The other H~II region complexes in NGC 4410
have H$\alpha$ luminosities ranging from 0.23 $-$ 1.0 $\times$ 
10$^{40}$ erg~s$^{-1}$.  
These lie near the top of the H~II region luminosity function 
in nearby spiral and irregular galaxies 
(Kennicutt, Edgar, \& Hodge 1989; Elmegreen \& Saltzer 1999).
The sizes of the HST subknots associated with these H~II regions, $\sim$90 pc,
are typical of those of OB associations in nearby spiral galaxies
(Bresolin, Kennicutt, \& Stetson 1996; Bresolin et al. 1998).

We can estimate how much of the emission-line luminosity arises from 
young stars by separately measuring the total emission excluding 
the two nuclei and the northwestern arc,
which are probably not ionized by young stars, and the emission from
the distinct H~II regions.
Excluding the two nuclei and the northwestern arc, the total observed L$_{H\alpha}$
for NGC 4410 is 1.5 $\times$ 10$^{41}$ erg~s$^{-1}$, uncorrected for
internal extinction.
The sum of the H$\alpha$ luminosities of the known H~II regions 
in NGC 4410 is 6.5 $\times$ 10$^{40}$ erg~s$^{-1}$.
The difference between these two luminosities is due to the extended ionized
gas outside of the northwestern arc; it is not yet clear whether
this is ionized by young stars or other mechanisms.
These two numbers define the range of the H$\alpha$ luminosity provided
by OB stars.  
Applying an extinction correction of A$_{\rm V}$ $\sim$ 1.2 (Table~\ref{Table4})
gives L$_{H\alpha}$ from star formation between 
1.6 $\times$ 10$^{41}$ erg s$^{-1}$
and 
3.6 $\times$ 10$^{41}$ erg s$^{-1}$.
These H$\alpha$ luminosities are typical
for nearby spiral and irregular galaxies
(Kennicutt 1983; Young et al. 1996).
Using the extended Miller-Scalo initial mass function (Kennicutt 1983),
the total star formation rate in NGC 4410 
is 1 $-$ 4 M$_\odot$ yr$^{-1}$.
This level of star formation relative to the blue luminosity in
NGC 4410 is relatively low compared to that in disk galaxies 
with similarly high blue luminosities.
The blue luminosity of NGC 4410A
is 5.7 $\times$ 10$^{10}$ L$_\odot$ (using B$_T$
from de Vaucouleurs et al. 1991 and M$_{B_{\sun}}$ = 5.48).
The L$_B$/L$_{H\alpha}$ ratio is 600 $-$ 1700, at the high
end of the range for spiral and irregular galaxies
(Young et al. 1996).
NGC 4410 is thus not a starburst: its current star formation 
rate in NGC~4410 is modest  
compared to its past star formation rate, as inferred from
the blue luminosity. NGC~4410A has an absolute blue magnitude
of -21.1, approximately that of M87. It is by far the most
luminous galaxy in the group, providing 25\% of its total
blue luminosity (Smith 2000). It may have come by its 
stellar content by a high rate of star formation in the past or by
stealing stars from other galaxies. The high blue luminosity of
NGC~4410A compared to its compansions and compared to its current
H$\alpha$ luminosity support the 
hypothesis that it is a spiral turning into an elliptical.

Star formation is probably contributing to the emission line 
luminosity seen in the nucleus of NGC~4410A in addition to that
inferred in the ring.
Summing only the flux from the NGC 4410A nucleus and the NW arc,
(in the arc, the ionization is probably not due to young stars),
$L_{H\alpha+[N+II]}/L_{rad}$ is $\sim 7$ times higher than
the typical value for non-compact, radio-selected radio galaxies.
The presence of this excess suggests that 
even in the central H$\alpha$+[N~II] source of NGC 4410A,
young stars may be contributing to the production of
H$\alpha$ emission.
As noted in \S\ref{opt_spec},
the spectra of the extended gas east and west of the nucleus
have line ratios intermediate between those expected for H~II regions
and LINERs, suggesting contributions from both classes of processes.

The NGC 4410 far-infrared luminosity of 3.9 $\times$ 10$^9$ L$\sun$
is moderate
compared to typical spiral galaxies (e.g. Young et al. 1996).
The L$_{FIR}$/L$_B$ ratio for this system is 0.068, which
is in the lower half of 
the range found for an infrared bright sample of spiral and 
irregular galaxies
studied by Young et al. (1989).
Therefore the
star formation rate implied by the far-infrared luminosity
is only somewhat smaller relative to the blue luminosity compared to
infrared-bright galaxies, which are probably forming stars more rapidly
than the typical spiral. 

Both the far infrared luminosity and the H$\alpha$ luminosity are
thought to be good tracers of star formation (e.g. Young et al. 1996).
The global L$_{FIR}$/L$_{H\alpha}$ ratio for NGC 4410, uncorrected
for extinction, is 67,
somewhat lower than typical for star-forming galaxies 
(e.g., Devereux \& Young 1990; Young et al. 1996).
However, using only the observed H$\alpha$ luminosity from young stars gives 
L$_{FIR}$/L$_{H\alpha}$ $\sim$ 100 $-$ 230,
consistent with ratios found for star forming
galaxies (Devereux \& Young 1990; Young et al. 1996, 
after correcting for different definitions of far-infrared
luminosity).
This similarity with star-forming galaxies suggests that massive stars
dominate the heating of dust in this system rather
than the active nuclei or shocks.  It also
implies that the H$\alpha$ extinction
is not high, consistent with what we find from the optical
spectroscopy.

Observations in the 2.6 mm CO (1 $-$ 0) line show that 
NGC 4410A is rich in molecular gas, with M$_{H_2}$ =
3.9 $\times$ 10$^9$ M$\sun$ (Smith 2000), assuming
the standard Galactic I$_{CO}$/N$_{H_2}$ conversion factor.
Comparing the H$\alpha$ luminosity from young stars to
this mass gives 
L$_{H\alpha}$/M$_{H_2}$ = 0.004 $-$ 0.009 L$\sun$/M$\sun$,
at the low end of the range for normal spirals and irregulars
(Young et al. 1996).   Thus NGC 4410 is not forming stars
at a particularly high rate compared to the available amount
of fuel for star formation, relative to normal spirals.
The L$_{FIR}$/M$_{H_2}$ ratio for NGC 4410A+B
is only $\sim$ 1 L$\sun$/M$\sun$ (Smith 2000), 
also implying a moderate star formation rate relative to the available
molecular gas, at the low end of the
range for normal galaxies (Young et al. 1996).
At present, CO fluxes of only a handful of radio galaxies
have been published to date (Mirabel et al. 1989; 
Mazzarella et al. 1993; Evans et al. 1999a,b; 
Lin et al. 2000).  
Many of these galaxies have 
L$_{FIR}$/M$_{H_2}$ ratios
higher than
that of NGC 4410A+B by a factor of $\sim$10. The
Mazzarella et al. (1993) galaxies, however,
were selected on the basis of their far-infrared brightnesses,
and have far-infrared luminosities an order of magnitude higher than that
of NGC 4410A, 
so are not an unbiased comparison sample for NGC 4410 in
terms of 
L$_{FIR}$/M$_{H_2}$. In the radio-selected survey of Lin et al. (2000),
only two of the 19 surveyed radio galaxies were detected above the
threshold $\sim 10^{8}~\rm{M}_\odot$. Compared to that sample, NGC~4410
has abundant molecular gas.

The CO line width in NGC 4410A is quite broad,
with FWHM $\sim$ 600 km s$^{-1}$ (Smith 2000).
Based on a comparison of the optical and CO velocities
(Smith 2000), 
about 60$\%$ of this
CO emission may be associated with the
H~II regions, while the remainder arises from the northwestern 
filamentary structure.
Thus for the eastern and western portions of the ring, 
L$_{H\alpha}$/M$_{H_2}$ $\sim$ 0.007 L$\sun$/M$\sun$
and $\sim$0.003 L$\sun$/M$\sun$, respectively. 
The value for the western portion should be considered
an upper limit in comparing with values derived
for star forming regions, as much of the ionization may be due to
shocks rather than young stars.

In contrast, the X-ray emission from NGC~4410 is probably dominated by
the AGN and an extended thermal source unrelated to the star 
formation.  
The X-ray emission from NGC 4410 in the ROSAT High Resolution
Imager (HRI) maps has been resolved into two
components: a point source with 
L$_{\rm X}$ = 2.7 $\times$ 10$^{41}$
erg~s$^{-1}$ associated with the nucleus of NGC 4410A,
and an extended (10$''$) halo with 
L$_{\rm X}$ = 1.3 $\times$ 10$^{41}$
erg~s$^{-1}$, offset towards Knot $\#$2 (Tsch\"oke et al. 1999).
Assuming the halo originates from the star forming regions,
it has L$_{\rm X}$/L$_{H\alpha}$ = 2.0.
For the center of NGC 4410A, L$_{H\alpha}$ = 4.2 $\times$ 10$^{40}$ erg
s$^{-1}$ (Tables~\ref{Table1} and~\ref{Table3}), so 
L$_{\rm X}$/L$_{H\alpha}$ = 6.4.
P\'erez-Olea \& Colina (1996) find that active nuclei
are distinguishable from starbursts by their
L$_{\rm X}$/L$_{H\alpha}$ ratio: starbursts have 
L$_{\rm X}$/L$_{H\alpha}$ $\le$ 0.6, while active galaxies
have higher ratios.
The two X-ray components in NGC 4410 both have
$L_x/L_{H\alpha}$ ratios higher than the starburst/AGN 
cutoff, implying that the halo component of the X-ray emission
may not be solely due
to star formation -- a hot ISM, like that found in giant ellipticals, 
could contribute to the X-ray luminosity.

\subsection{Comparison to Other Radio Galaxies}

There have been numerous narrowband optical imaging studies
of radio galaxies in the past, and the distribution
of ionized gas has been mapped in more than 100 radio
galaxies (Hansen, N$\o$rggard-Nielsen, \& J$\o$rgensen
1987;
Baum et al. 1988;
Morganti, Ulrich, \& Tadhunter 1992;
McCarthy, Spinrad, \& van Breugel 1995; Hes, Barthel, \&
Fosbury 1996).
In most of these systems, the ionization of the extended
gas is believed
to be dominated by either photoionization by the active nucleus
or shock ionization, rather than young stars (Robinson
et al. 1987); such inferences, however, are usually made on the
basis of composite spectra. 
In NGC 4410, $\sim$25$\%$ of the total H$\alpha$+[N~II]
originates from the discrete H~II regions, $\sim$60$\%$ 
comes from the two nuclei and the northwestern arc,
and 15$\%$ is due to diffuse gas outside of these regions.
Therefore, even in NGC 4410, young stars do not dominate the ionization.
The relative proximity of NGC 4410 to us, however, enables us
to spatially resolve structures indistinguishable at higher
redshifts, in particular, to separate H~II regions from gas
ionized by other means. In this section, we compare the global properties
of NGC 4410 to those of radio galaxies, and, since we resolve H~II
regions in NGC 4410A, we discuss what these observations mean for 
interpretation of emission
line regions in radio galaxies.

NGC 4410 appears to have a high emission-line luminosity 
(L$_{H\alpha+[NII]}$) 
compared to most radio galaxies.  
For radio-selected radio galaxies, there is
a strong correlation between the H$\alpha$+[N~II]
luminosity and the total radio luminosity L$_{rad}$.  
This correlation
holds not just for high luminosity radio galaxies 
(Baum \& Heckman 1989), but also for
low luminosity systems (Morganti, Ulrich \& Tadhunter 1992), like
NGC4410.  Morganti et al. (1992) find a relationship between
$L_{H\alpha+[N+II]}$ and $L_{rad}$ for radio sources,
excluding compact (size $\le$10 kpc) radio sources,
over five orders of magnitude in radio
luminosity. Morganti et al. (1992) find $\log 
L_{H\alpha+[N+II]} = 0.73 \pm 0.12 \log L_{rad} + 9.93 \pm 4.20$.

The  L$_{H\alpha+[N+II]}$/L$_{rad}$ ratio (but not the
linear radio size) of  NGC~4410A is similar to the
Morganti et al. (1992)  
compact radio sources.  The
compact sources in the Morganti et al. (1992) sample have much higher
$L_{H\alpha+[N+II]}/L_{rad}$ ratios than the rest of the Morganti sources,
up to 20 times higher. Similarly, since 
NGC 4410A has a total radio luminosity of 
$1.3 \times 10^{41}$ erg s$^{-1}$
(Hummel et al. 1986), $L_{H\alpha}$ for NGC~4410A is
30 times higher than predicted by the Morganti et al. (1992) relation 
for non-compact radio sources.
Morganti et al. (1992)
suggest that star formation contributes significantly to the 
ionization of gas in the compact sources. Star formation 
contributes significantly in NGC~4410A as well. 
However, the radio source in NGC 4410A is much larger (280 kpc) than
the compact ($\le$10 kpc) radio galaxies in the Morganti et al. (1992)
sample.

The far infrared properties of NGC~4410 are consistent with those
of a radio galaxy with dust heated by star formation. From
the low ratio of far-infrared to radio luminosity (\S\ref{Intro}), the 
radio emission is clearly dominated by an active nucleus. However,
strong far-infrared emission is not unusual in radio galaxies.
In the Golombek, Miley, \& Neugebauer (1988) 
study of 131 radio galaxies, 44$\%$ were detected,
half with far-infrared luminosities greater than that of NGC 4410.
Similar results were found by Impey, Wynn-Williams, \& Becklin (1990) 
for a radio-selected sample.
In NGC 4410, as well as in many of the other radio galaxies detected
by IRAS, this far-infrared emission is due to interstellar dust,
not direct synchrotron radiation from the active nucleus.
This is demonstrated for NGC 4410 in Figure~\ref{fig9}, where we plot
the global spectral energy distribution of NGC 4410A+B.
A clear far-infrared excess is seen above the power-law
radio continuum.
Such an excess is also found in most of the radio galaxies
surveyed by Impey et al. (1990).
In NGC 4410, the 
60 $\mu$m/100 $\mu$m flux density ratio is $\sim$0.5 
(Mazzarella et al. 1991), within the range found for star forming galaxies
(Helou 1986), supporting the idea the far infrared emission could be
produced by dust heated by star formation.
To distinguish between dust heated by an active nucleus
and star formation, the 25 $\mu$m/60 $\mu$m ratio is often
used (e.g., de Grijp et al. 1985; Miley, Neugebauer, \& Soifer 1985);
an enhanced F$_{25~{\mu}m}$/F$_{{60~\mu}m}$ ratio
signals warmer dust than expect from star formation alone.
Based on the IRAS data for NGC 4410, which provides
only a $3\sigma$ upper limit at
25 $\mu$m/60 $\mu$m of 0.59, we cannot rule
out the possibility that emission from hot dust near   
the active nucleus may contribute to some the
far infrared emission. 

Whereas the detection of strong far-infrared
emission from numerous radio galaxies indicates that many 
have reasonably high star formation rates,
the direct observations of luminous H~II regions in radio
galaxies via narrowband optical imaging is still uncommon.
One of the few other radio galaxies with H~II regions clearly
distinguishable
in H$\alpha$+[N~II] maps 
is the peculiar galaxy Centaurus A, which has an inclined
dusty disk with 
numerous H~II regions
(Hodge \& Kennicutt 1983;
Bland, Taylor, \& Atherton 1987).
The far-infrared luminosity of Cen A is 5 $\times$ 10$^9$ L$\sun$
(Rice et al. 1988, using a distance of 3.5 Mpc from
Hui et al. (1993)), similar to that of NGC 4410A+B.
Most of this far-infrared
emission arises from the Cen A disk rather
than its nucleus (Joy et al. 1988).

Cygnus A 
(Stockton, Ridgway, \& Lilly 1994;
Jackson, Tadhunter, \& Sparks 1998) 
and Hydra A (Melnick, Gopal-Krishna,
\& Terlevich 1997)
also appear to have some circumnuclear star formation,
however, the star formation in these systems
is much less prominent than that in
Cen A or NGC 4410.
Fornax A 
(Mackie \& Fabbiano 1998)
and PKS 0349-27
(Grimberg,
Sadler, \& Simkin 1999)
show little
evidence for the presence of OB stars.

Star formation in a radio galaxy may be triggered
by density waves in a disk, as is presumably occurring
in the Cen A disk.
Alternatively, star formation may be induced
by the radio lobe impacting the interstellar medium,
which may be the case for Minkowski's object,
a star formation region near the radio galaxy NGC 541
(van Breugel
et al. 1985).   This object has an H$\alpha$ luminosity
of 3 $\times$ 10$^{40}$ erg s$^{-1}$, similar to that of
the most luminous H~II region in NGC 4410.
Another example of jet-induced star formation
may have occurred in 3C 285; like Minkowski's object, the observed H~II
region in this system has L$_{H\alpha}$ = 3 $\times$ 10$^{40}$ erg~s$^{-1}$
(van
Breugel \& Dey 1993).
Jet-induced star formation may also be occurring in Cen A,
outside of the main disk.
Blue stars and H~II regions
have been found at the boundary of one of the radio lobes and an HI cloud 
(Blanco et al. 1975;
Graham \& Price 1981; Graham 1998).
The Cen~A H~II regions, however, have H$\alpha$ luminosities
of only $\sim$3 $\times$ 10$^{37}$ L$\sun$, much lower than those
found in NGC 4410.

\subsection{What Triggered the Star Formation in NGC 4410?}

The peculiar star formation morphology of NGC 4410 has several
possible origins.   Star formation may have been triggered
by an expanding density wave, as in classical ring galaxies
(e.g., Lynds \& Toomre 1976; Theys
\& Spiegel 1977).
The existence of a underlying 
stellar counterpart to the ring
supports this hypothesis; this ring is not solely
new stars.  The presence of dust on both sides of the ring,
in somewhat continuous structures, also supports this idea.
If this is indeed a continuous ring, however,
it is somewhat distorted.
The ring has a position angle of $-$30$^{\circ}$ east of north,
and an axial ratio of $\sim$0.65 (Figures~\ref{fig2} and \ref{fig4}).
If this structure is an inherently round ring, the most
blueshifted and most redshifted gas would be expected
to lie along the major axis.  Instead, the line of nodes
runs roughly east-west, with the most redshifted gas in
the northwestern arc, and the most blueshifted material
in Knot $\#$1
(Figure~\ref{fig7}).

If the ring was indeed caused by a head-on collision,
the most likely collider is NGC 4410B.
The blue luminosity of NGC 4410B is 1/6th
that of
NGC 4410A (de Vaucouleurs
et al. 1991), implying an unequal mass collision. 
Rings formed by such collisions typically have lifetimes of
$\sim$10$^8$ years (Theys $\&$ Spiegel 1977;
Gerber, Lamb, $\&$ Balsara 1996),
consistent with estimates made for NGC 4410A+B based on 
pair separation and ring size.
Neglecting unknown projection
effects, the current separation and velocity
difference of the two
galactic nuclei, 9.5 kpc and 60 km s$^{-1}$, imply 
a time since collision of 1.6 $\times$ 10$^8$ years.
With a projected ring radius of 8$''$ $-$ 15$''$ (4 $-$ 7 kpc),
assuming an expansion velocity typical
of ring galaxies of 50 $-$ 100 km s$^{-1}$
(e.g., Struck-Marcell $\&$ Higdon 1993; Gerber et al. 1996), we estimate
a collision timescale of 0.4 $-$ 1.4
$\times$ 10$^8$ years for NGC 4410A, also consistent with theoretical
models of ring galaxies.

If NGC 4410A is a classical ring galaxy, however, the 
observation of line ratios in some of the
features consistent with shock ionization may be surprising.
The line ratios in ring galaxies are generally typical of H~II regions,
not shocks (Jeske 1986; Bransford et al. 1998). 

Some evidence for an interaction in the group 
is visible in the X-ray data, in a fainter third X-ray component.
In the lower resolution Position Sensitive Proportional Counter (PSPC)
map, a faint tail-like structure containing $\sim$10$\%$
of the total X-ray flux extends 2$'$ (56 kpc) to the east
coincident with the stellar bridge connecting NGC 4410A+B
and NGC 4410C (Tsch\"oke et al. 1999; Smith 2000). Such
a structure may arise in an interaction between NGC~4410C and
its group companions.

Alternatively, star formation
may have been triggered by the radio lobe impacting the interstellar
medium, as suggested for Minkowski's object
(van Breugel et al. 1985).
The fainter radio continuum lobe of NGC 4410 extends to the
northwest towards the ring (Smith 2000), 
however, at present the resolution of the
available radio continuum maps is not high enough to 
determine its spatial association with the H~II regions. High resolution
radio data would be useful to reveal the answer to this question.
The spectroscopic evidence for shocks or at least
compressed gas in the northwestern arc,
however, supports
this alternative hypothesis.

A third possibility is that the ring is the remnant of
a third galaxy which has been torn apart by a collision.
In this case, Knot $\#$2 may be a third nucleus.  The relatively
narrow
optical line widths of this source (Section 3.4),
however, argues against this interpretation.

\subsection{The Nature of the Optical Knots in the Southeastern Tail}

In the southeastern NGC 4410 lobe,
a prominent radio `hot spot' is found,
coincident
with an optical knot (Hummel et al. 1986). 
Hummel et al. (1986) suggested that this optical
structure may be associated with the radio knot in some manner.
It is possible that this source is 
a `hot spot' emitting optical synchrotron emission,
as is believed to be the case for the optical knots studied by 
Keel \& Martini (1995),
O'Dea et al. (1999),
and 
L\"ahteenm\"aki \& Valtaoja (1999).
The 4.9 and 1.5 GHz flux densities of the NGC 4410 knot are
2.5 mJy and 4.4 mJy, respectively, with a spectral
index $\alpha$ (F$_{\nu}$ $\propto$ $\nu$$^{\alpha}$)
of $-$0.47.
The optical counterpart has an R magnitude of $\sim$19 (\S3.2),
giving an optical/4.9 GHz
spectral index of $\sim$$-$0.31, similar to that
in the radio. The estimated index supports the idea
that the optical emission is due to synchrotron emission.
However, 
the NGC 4410 knot is unresolved in our narrowband red image
(FWHM $\le$0\farcs9
$\sim$ 400 pc), while the radio continuum knot has a size
of $\sim$15$''$ (7 kpc) (Hummel et al. 1986),
arguing against an association.
For comparison, the optical hot spots studied by
L\"ahteenm\"aki \& Valtaoja (1999)
have linear sizes between 1 and 7 kpc.

Alternatively, this knot may be an H~II region, either
caused by the interaction or induced by 
the pressure of the jet impacting
the 
interstellar medium.
Evidence for such jet-induced star formation has been found
in a number of systems
(e.g., de Young 1981;
van Breugel et al. 1985;
van Bruegel \& Dey 1993; 
Graham 1998).
This optical knot, however, was not detected in our
H$\alpha$+[N~II] map,
giving
L$_{H\alpha}$
$\le$ 3.5 $\times$ 10$^{38}$ erg s$^{-1}$. 
This is lower than
the luminosities of the H~II regions in the tails of the Antennae galaxy
(Mirabel, Dottori, \& Lutz 1992) and the Mice
pair (Hibbard \& van Gorkom 1996), 
and 
lower than the luminosity of Minkowski's object.
However, this upper limit is 
consistent with the H$\alpha$ luminosities of the H~II regions
in the tails of NGC 2782 
(Smith et al. 1999), Arp 295, NGC 520, and NGC 3621
(Hibbard \& van Gorkom 1996), as well as
the possible jet-induced H~II regions in Cen A
studied by Graham (1998).
Thus, the lack of observed H$\alpha$ emission associated with
the northern knot in the NGC 4410 tail does not rule out 
on-going star formation.

A third possibility is that these knots may be old stellar clusters.  
If they are at the distance of NGC 4410, they have M$_{\rm R}$ $\sim$
$-$16,
similar to the most luminous of the so-called `super star
clusters' found in the inner regions of interacting galaxies by
HST (e.g., Holtzman et al. 1992).
Finally, these sources could be background or foreground objects.
These last possibilities could be resolved with optical colors and/or
spectroscopy.

\subsection{The NW Arc}

The broad H$\alpha$ line and the strong forbidden lines in the
NW arc show that it is not ionized by standard H~II regions
(Section 3.4), particularly evident from the high [OI]6300\AA / H$\alpha$ 
ratio.  
Line ratios in this range are typical of AGN (e.g. Veron-Cetty \& Veron 2000), 
cooling flows 
(Heckman et al. 1989; Voit \& Donahue 1995), and 
gas that has been shock-ionized (e.g. Shull \& McKee 1979; 
Dopita \& Sutherland 1996). The cooling flow spectra
can be explained by a composite of hot star photoionization and some
other source of heat (Voit \& Donahue 1995); 
very slow shocks could contribute, but Voit \&
Donahue (1995) and Donahue et al. (2000) argue that such shocks are
very inefficient at producing H$\alpha$.
Extended emission to one side
of the nucleus could be
produced by the asymmetric production of radiation from the 
AGN, as in the cone structures observed
in Seyferts (e.g. Capetti et al.  1994; Wilson et al. 1989). The emission
is similar to the line emission produced by collisional debris as 
seen in NGC~4438, a spiral galaxy which has apparently recently
interacted with a companion S0 galaxy (Kenney et al. 1995).
Extensive plasma diagnostics to sort out the various 
possibilities are not possible with the spectrum in 
hand, lacking the detection of [OIII] and other fainter diagnostic
lines.

Based on the ROSAT X-ray image of this system, we can rule
out a cooling flow as the dominant source of heat 
for the NW arc.
The thermal X-ray emission from the NGC 4410A+B vicinity is $\leq 1.3 \times
10^{41} h_{75}^{-2}$ erg s$^{-1}$ (Tsch\"oke et al. 1999); this value is only 
an upper limit
to the luminosity arising from cooling flow gas, since some of the X-ray 
radiation
must originate from X-ray binaries in the star formation regions (Section 4.1).
For comparison, the H$\alpha$ luminosity of the NW arc is
2 $\times$ 10$^{40} h_{75}^{-2}$ erg s$^{-1}$ (Tables~\ref{Table1} and 
\ref{Table3}).
Therefore L$_{H\alpha}$/L$_X$ $\ge$ 0.15, 
much higher than
typical cooling flow clusters, where L$_{H\alpha}$/L$_X$ 
is less than 0.02, even 
when the X-ray luminosity from the cooling flow is computed in a region
local to the optical filaments (Heckman et al. 1989).

In other radio galaxies, there have been numerous observations of
extended ionized gas exterior to the nuclear regions.
As discussed in \S4.2, in some cases, H~II regions have been
found in other radio galaxies.  In most cases, however, the extended ionized gas found
in other radio galaxies has line
ratios consistent with shock ionization (Robinson et al. 1987),
similar to those found in the NW arc of NGC 4410.

\section{Conclusions}

NGC 4410A is the host of both a radio-loud active nucleus
and on-going star formation.
The star formation is in the form of extremely luminous
H~II regions aligned in an arc along the eastern portion
of a ring-like stellar structure surrounding the active nucleus.
The western portion
of this ring contains filamentary ionized gas with optical
line ratios implying shock ionization. NGC~4410A is thus a 
nearby example of an AGN with spatially-resolved 
evidence for both star formation and shocks in an extended
emission line region. Such an object lends credence to the
hypothesis that the emission line regions associated with radio
galaxies and the central galaxies in cluster cooling flows 
might also result from the effects of more than one phenomenon, with
the implication that  
no single physical process can explain all of the observations.
NGC 4410A could be an example of a disk galaxy on its way to
becoming an elliptical, via interactions with its group companions.

\acknowledgements

We thank James Webb for help with the SARA observations. We are
grateful to Cathy Imhoff (MAST) for assistance with the NEWSIPS
package and discussions regarding the IUE data. 
This research has made use of the NASA/IPAC Extragalactic
Database (NED) which is operated by the Jet Propulsion
Laboratory under contract with NASA. Some of the data presented
in this paper were obtained from the Multimission Archive at
the Space Telescope Science Institute (MAST). STScI is operated
by the Association for Universities for Research in Astronomy,
Inc. under NASA contract NAS5-26555. Support for MAST for non-HST
data is provided by the NASA Office of Space Science via
grant NAG5-7584 and by other grants and contracts. We would
also like to acknowledge the use of the Double Spectrograph at
the Palomar Observatory (5-meter), then under joint support by the
Observatories of the Carnegie Institution of Washington and the
California Institute of Technology. We also obtained data from
facilities at the Kitt Peak National Observatory, which is
funded by the National Science Foundation and operated by
the Association of Universities for Research in Astronomy, Inc.
B. J. S. acknowledges support from NASA grant AR-08374.01-97A from
the Space Telescope Science Institute.

\eject
\clearpage

\begin{figure}
\plotone{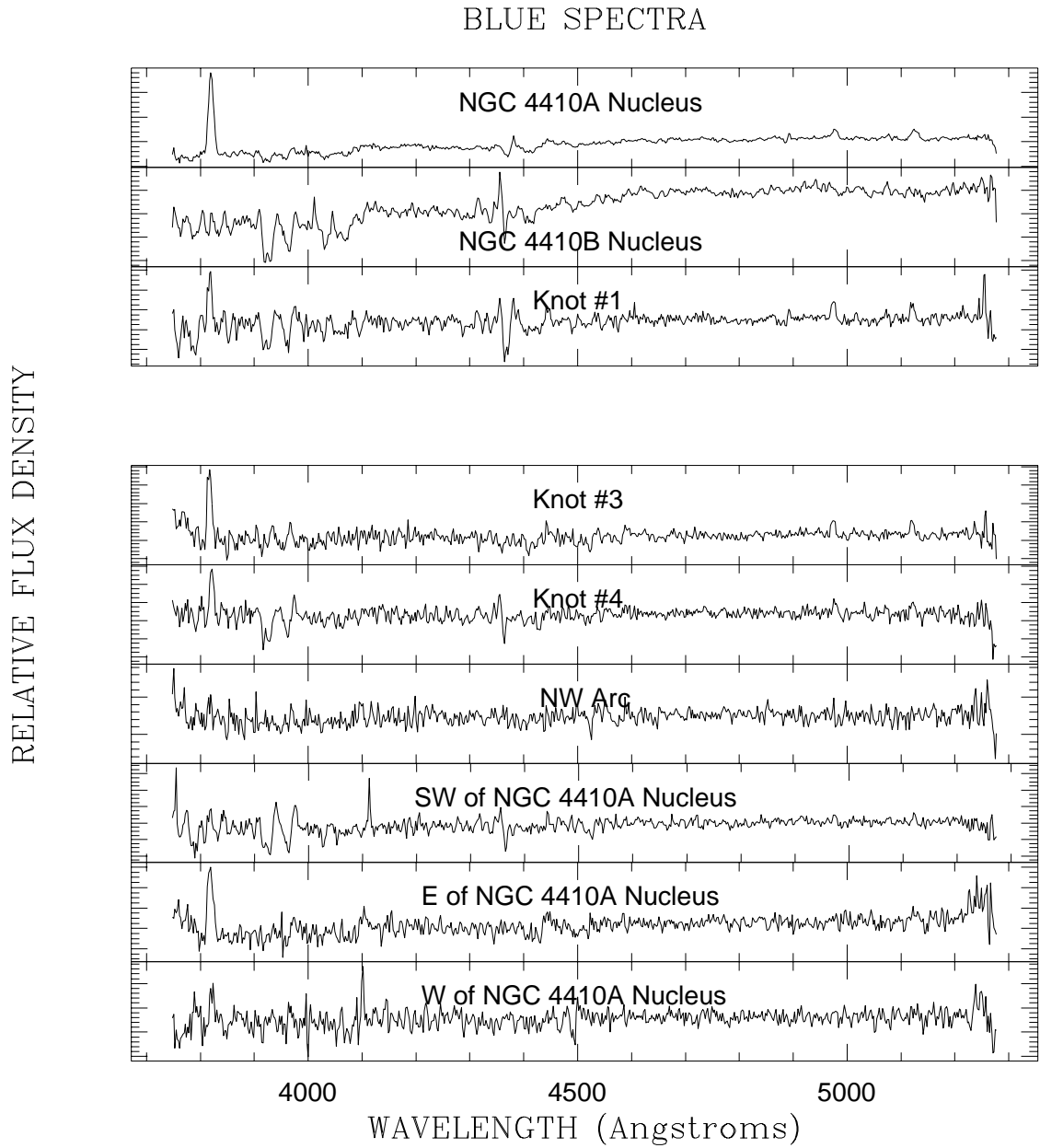}
\caption{A: The blue optical spectra for the 10 selected positions
in NGC 4410.}
\end{figure}
\setcounter{figure}{5}
\begin{figure}
\plotone{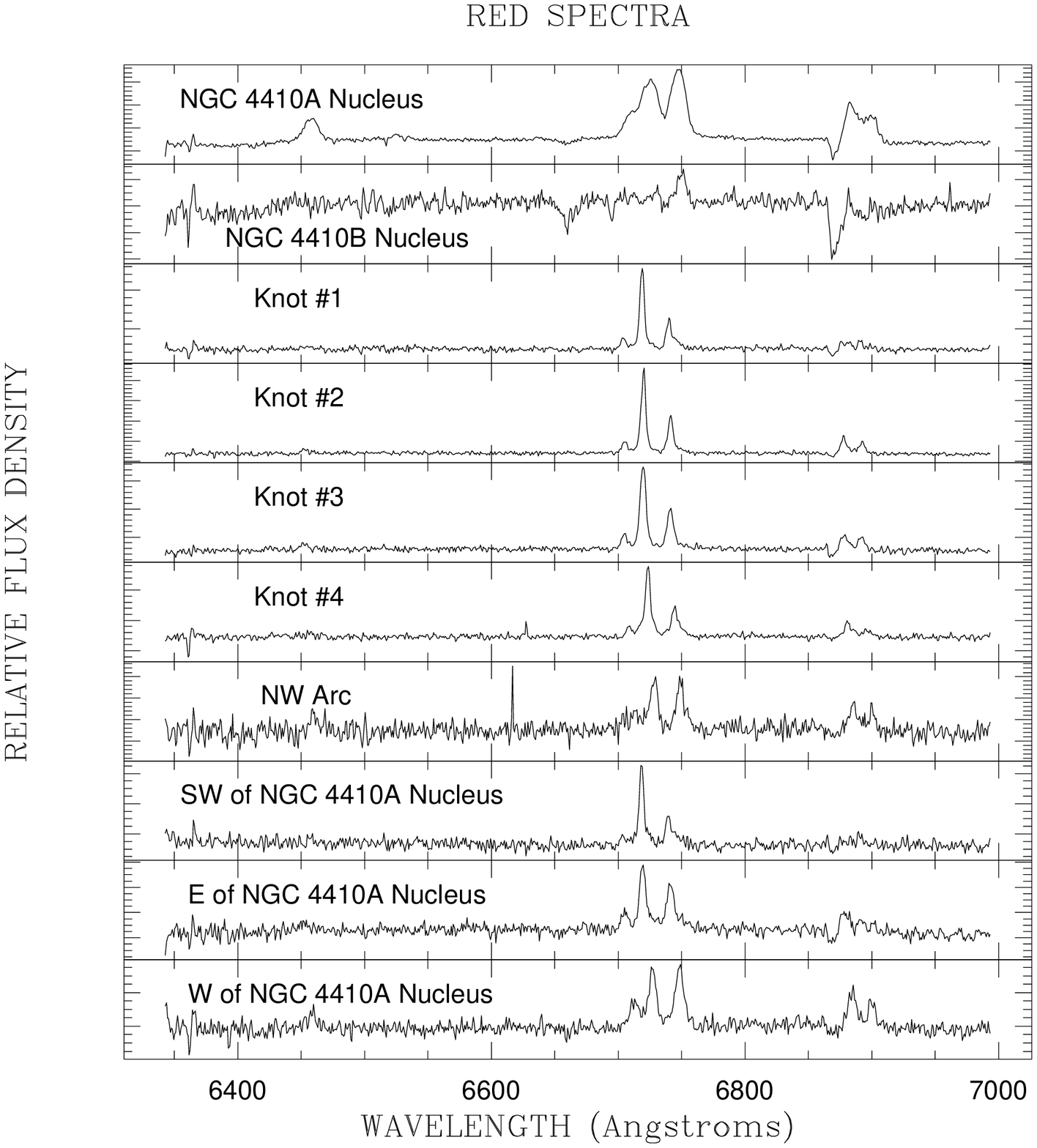}
 \caption{B: The red spectra.
Multiple observations have been averaged.
In the blue spectrum from Observation \#4, because of the poorer camera
optics on the blue side of the Double Spectrograph,  Knot \#2 was spatially unresolved
from the nucleus. Therefore we do not provide a blue spectrum for this source.
\label{fig6}}
\end{figure}

\begin{figure}
\plotone{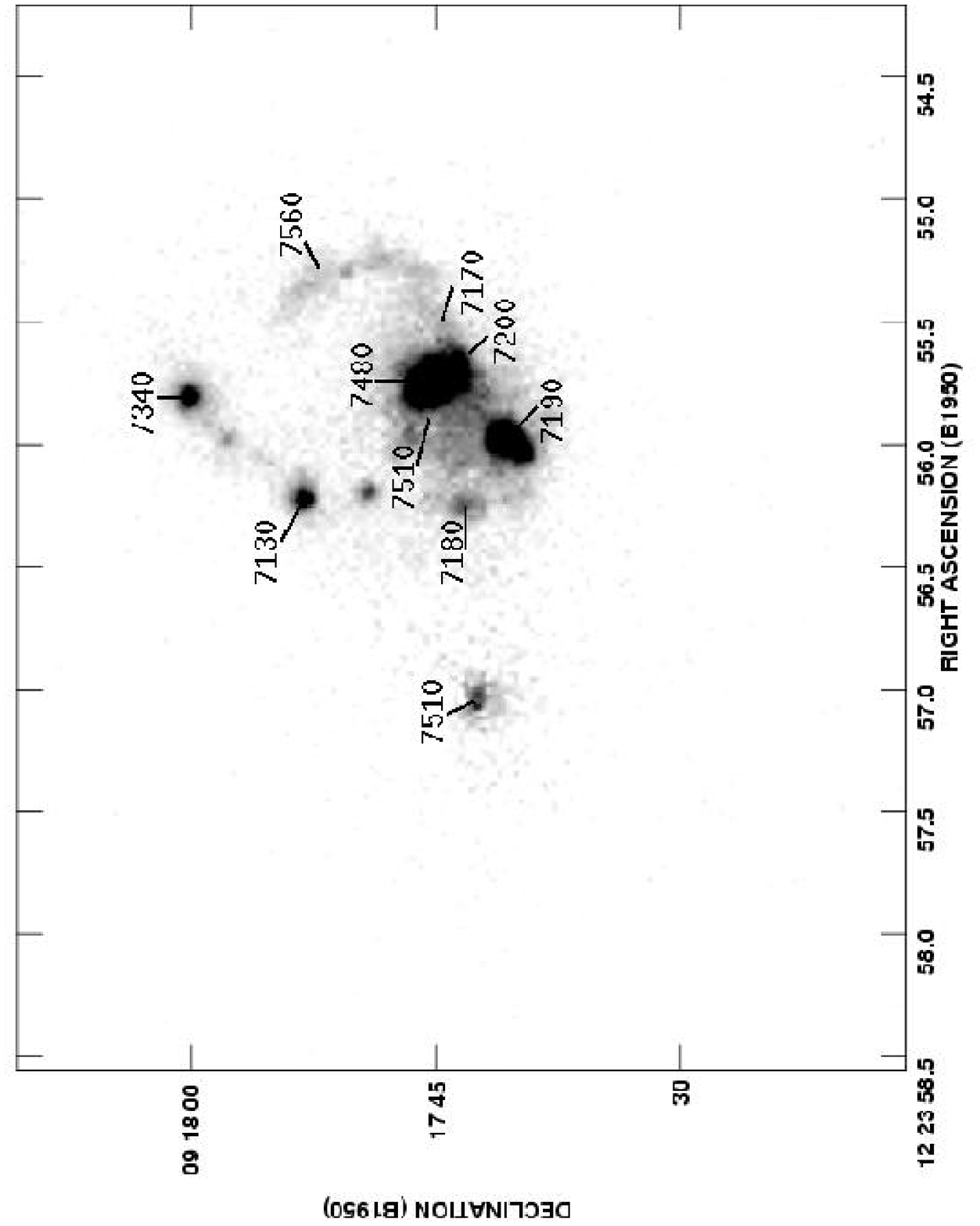}
 \caption{
The optical velocities marked on the H$\alpha$+[N~II]
image.
\label{fig7}}
\end{figure}

\begin{figure}
\plotone{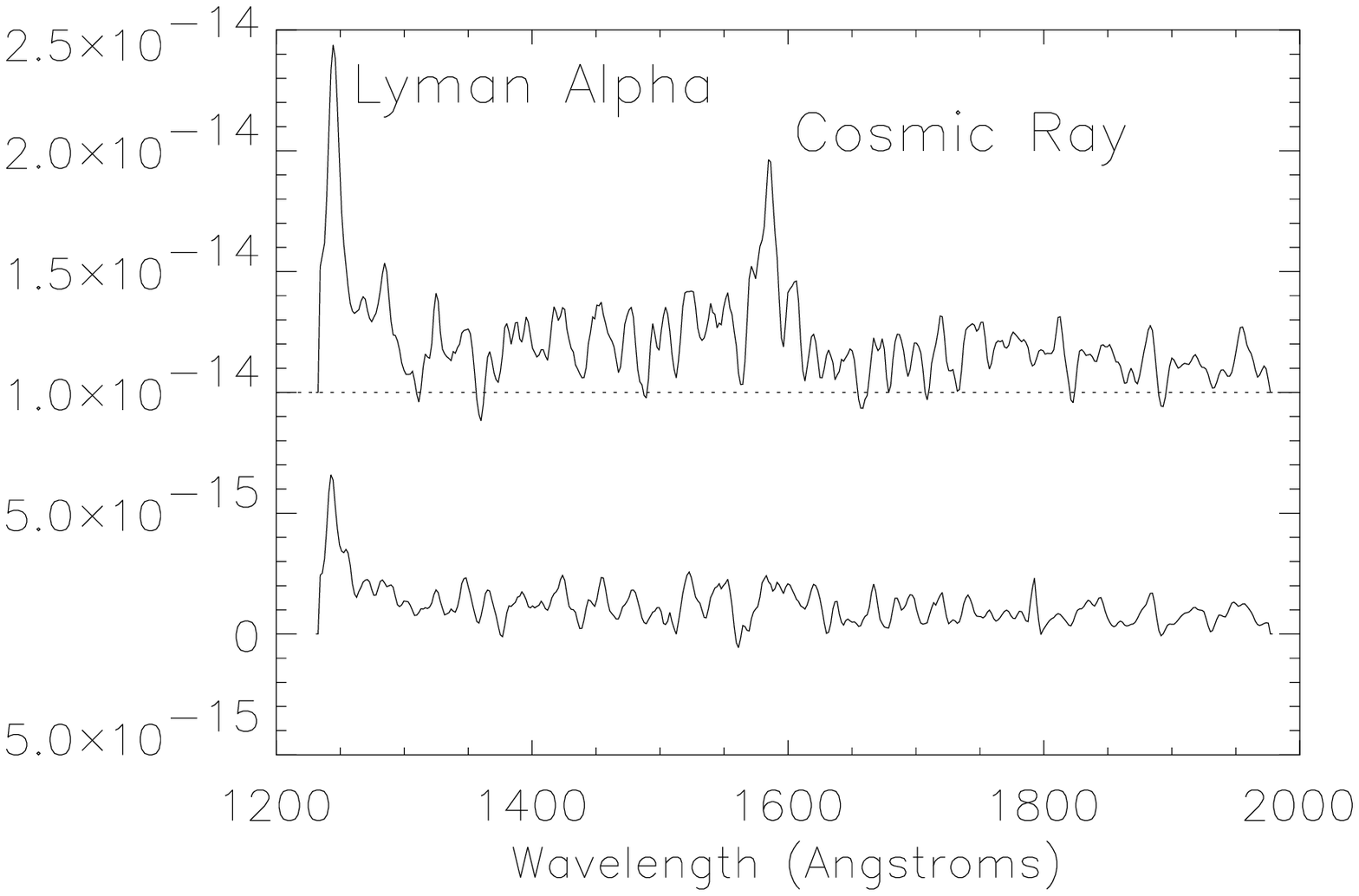}
 \caption{
The smoothed IUE SWP spectra for NGC 4410A.
The top spectrum was obtained by reprocessing the data with 
the extended source software in NEWSIPS.  The lower 
spectrum was extracted using the standard NEWSIPS point source
algorithm. The y-axis is flux, in erg s$^{-1}$ cm$^{-2}$ \AA$^{-1}$
units. For display purposes,
the upper spectra has been offset by 
$10^{-14}$ erg s$^{-1}$
cm$^{-2}$ ${\rm \AA}$$^{-1}$, marked as a baseline on the figure.
The variations in the continuum are at the level of the noise. The 
only detected feature is the Lyman-$\alpha$ emission line in both the
extended spectrum and the point-source spectrum. The feature near
1570\AA~ in the extended source spectrum was visually identified as
a cosmic ray hit.
\label{fig8}}
\end{figure}

\begin{figure}
\plotone{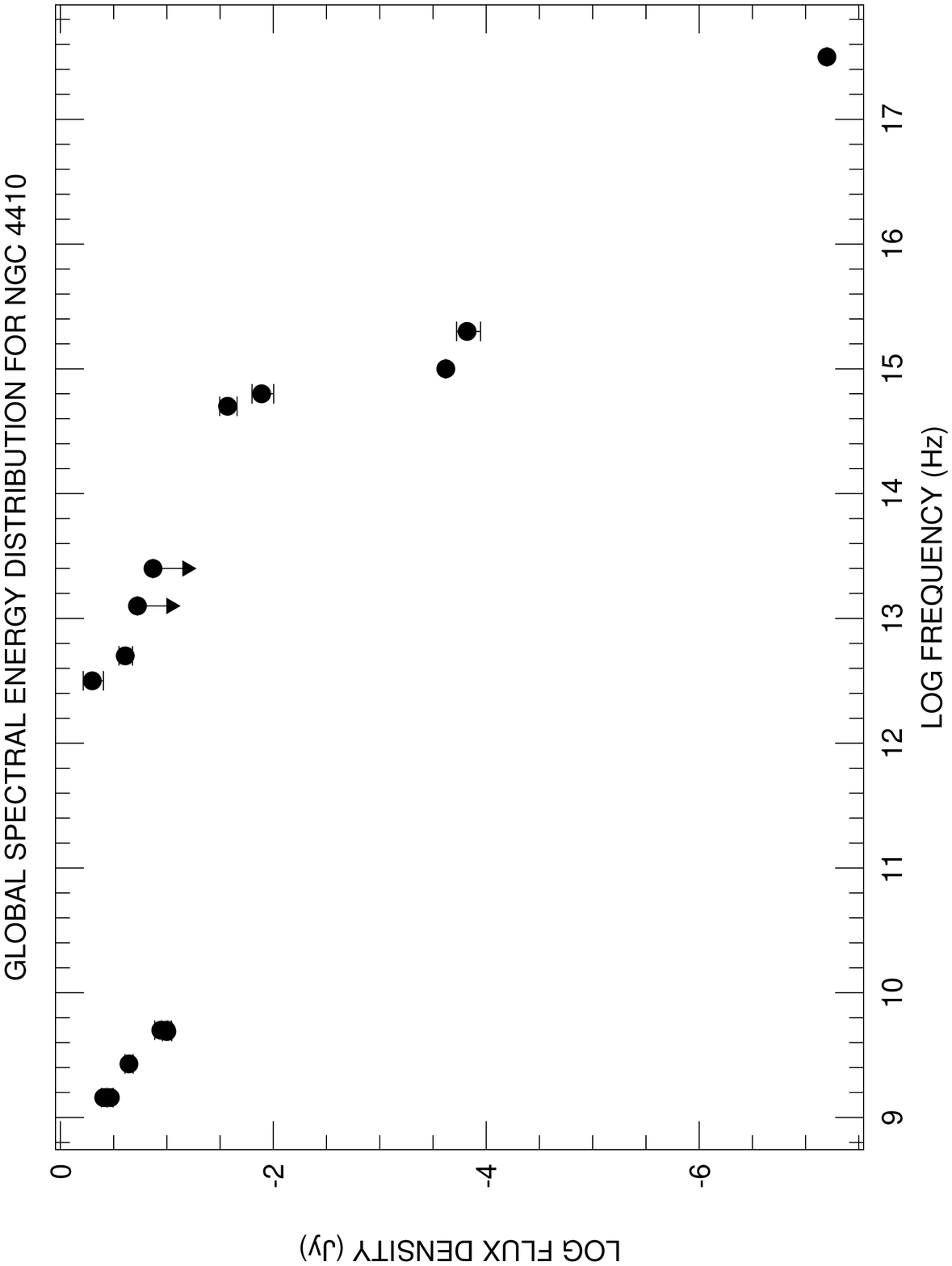}
 \caption{
The global spectral energy distribution for NGC 4410.
The radio data are from Hummel et al. (1986), the far-infrared
from 
Mazzarella et al. (1991),
the optical measurements from de Vaucouleurs et al. (1991),
the ultraviolet data from this work, and the X-ray points from
Tsch\"oke et al. (1999).
\label{fig9}}
\end{figure}

\clearpage

\begin{table}
   \caption{Nuclei and H~II Regions in the NGC 4410A+B System \label{Table1}}
   \begin{tabular}{lccccccccccccccc} \hline
       \multicolumn{1}{c}{Name}&
\multicolumn{6}{c}{Position}
&\multicolumn{1}{c}{L$_{H{\alpha}+[N~II]}$$^a$}
&\multicolumn{1}{c}{Velocity$^b$}
\\
       \multicolumn{1}{c}{}&
\multicolumn{3}{c}{R.A. (1950)}&
\multicolumn{3}{c}{Dec. (1950)}&
\multicolumn{1}{c}{(erg s$^{-1}$)}& 
\multicolumn{1}{c}{(km s$^{-1}$)}& \\
\hline
NGC 4410A Nucleus&12&23&55.78&9&17&45&9.0 $\times$ 10$^{40}$&7440\\
NGC 4410B Nucleus&12&23&57.03&9&17&43&1.5 $\times$ 10$^{40}$&7500\\
Knot $\#$1&12&23&56.24&9&17&53&1.2 $\times$ 10$^{40}$&7140\\
Knot $\#$2&12&23&55.96&9&17&41&5.0 $\times$ 10$^{40}$&7150\\
Knot $\#$3&12&23&56.26&9&17&43&8.8 $\times$ 10$^{39}$&7140\\
Knot $\#$4&12&23&55.81&9&18&0&1.5 $\times$ 10$^{40}$&7300\\
Knot $\#$5&12&23&55.99&9&17&58&3.5 $\times$ 10$^{39}$&\\
Knot $\#$6&12&23&56.19&9&17&49&6.1 $\times$ 10$^{39}$&\\
NW Arc&&&&&&&3.8 $\times$ 10$^{40}$&7490\\
\tablenotetext{a}{All luminosities were 
derived assuming a distance of 97 Mpc (H$_o$ = 75 km s$^{-1}$
Mpc$^{-1}$). No corrections for internal extinction were applied.
}
\tablenotetext{b}{Heliocentric velocity from the optical spectroscopy presented in this
paper.  The uncertainty
on these values is $\sim$100~km~s$^{-1}$.}
\end{tabular}
\end{table}

\begin{table} 
\caption{Slit Positions for Optical Spectroscopy\label{Table2}} 
   \begin{tabular}{ccccccccccclclcrcccc} \hline
      \multicolumn{1}{c}{Observation}&
\multicolumn{1}{c}{Center Position}
&\multicolumn{1}{c}{Position}&
\multicolumn{1}{c}{Additional}\\
      \multicolumn{1}{c}{}&
\multicolumn{1}{c}{of Slit}
&\multicolumn{1}{c}{Angle}&
\multicolumn{1}{c}{Objects Observed}\\
\hline
Observation 1 &NGC 4410A Nucleus&25$^{\circ}$&Knot $\#$1, Emission SW of A\\
Observation 2 &Knot $\#$1&135$^{\circ}$&NGC 4410B Nucleus, Knot $\#$4\\
Observation 3 &Knot $\#$1&0$^{\circ}$&Knot $\#$3\\
Observation 4 &NGC 4410A Nucleus&310$^{\circ}$&Knot $\#$2,
NW Arc\\
Observation 5 &NGC 4410A Nucleus&270$^{\circ}$&Emission E and W of A\\
\end{tabular}
\end{table}

\begin{table}
\caption{Optical Line Ratios\tablenotemark{a} \label{Table3}}
\begin{tabular}{llcccccccccccc}\hline
ID/obs& Obs\tablenotemark{b}& $\frac{[N~II]6584}{H\alpha}$& $\frac{[S~II]6717}{H\alpha}$& 
$\frac{[S~II]6717}{[S~II]6731}$&$\frac{[O~I]6300}{H\alpha}$&
$\frac{[O~III]5007}{[O~II]3727}$\tablenotemark{c,d} &$\frac{H\alpha}{H\beta}$\tablenotemark{d} &$\frac{[O~III]5007}{H\beta}$ \\
\hline
N4410A&1&0.97$\pm$0.01&0.44$\pm$0.01&1.37$\pm$0.09&0.28$\pm$0.01&0.16$\pm$0.01&2.90$\pm$0.09&0.98$\pm$0.04&\\
&4&1.23$\pm$0.01&0.56$\pm$0.01&1.38$\pm$0.05&0.39$\pm$0.01&0.15$\pm$0.01&4.20$\pm$0.13&1.45$\pm$0.05&\\
&5&1.19$\pm$0.01&0.51$\pm$0.01&1.27$\pm$0.05&0.39$\pm$0.01&0.13$\pm$0.01&5.00$\pm$0.17&1.35$\pm$0.06& \\
N4410B&2 & 2.83$\pm$0.63    & $\le$0.50   & $---$     & $\le$0.50   & $---$     & $---$     & $---$     & \\
Knot $\#$1&1 & 0.39$\pm$0.06    & 0.07$\pm$0.06    & 0.58$\pm$1.56    & 0.06$\pm$0.06    & 1.24$\pm$0.19    & 1.57$\pm$0.18    & 1.00$\pm$0.13    & \\
&2 & 0.51$\pm$0.05    & 0.11$\pm$0.05    & 1.99$\pm$2.87    & 0.14$\pm$0.05    & 0.37$\pm$0.04    & 2.39$\pm$0.30    & 1.26$\pm$0.18    & \\
&3 & 0.37$\pm$0.02    & 0.18$\pm$0.02    & 1.62$\pm$0.45    & $\le$0.04    & 0.17$\pm$0.02    & 3.55$\pm$0.21    & 0.57$\pm$0.07    & \\
Knot $\#$2$^c$&4 & 0.47$\pm$0.02    & 0.25$\pm$0.02    & 1.32$\pm$0.23    & 0.07$\pm$0.02    & $---$     & $---$     & $---$     & \\
Knot $\#$3&3 & 0.50$\pm$0.01    & 0.20$\pm$0.01    & 1.11$\pm$0.13    & 0.09$\pm$0.01    & 0.16$\pm$0.01    & 5.13$\pm$0.22    & 0.92$\pm$0.06    & \\
Knot $\#$4&2 & 0.47$\pm$0.02    & 0.25$\pm$0.02    & 1.85$\pm$0.51    & 0.04$\pm$0.02    & 0.23$\pm$0.03    & 6.55$\pm$0.94    & 1.00$\pm$0.20    & \\
NW Arc&4 & 0.97$\pm$0.10    & 0.48$\pm$0.08    & 1.38$\pm$0.58    & 0.27$\pm$0.07    & $---$     & $---$     & $---$     & \\
SW of A&1 & 0.66$\pm$0.05    & 0.32$\pm$0.04    & 1.25$\pm$0.45    & 0.20$\pm$0.04    & $---$     & 6.80$\pm$1.88     & 1.22$\pm$0.43     & \\
E of A&5 & 0.77$\pm$0.05    & 0.33$\pm$0.04    & 2.05$\pm$0.85    & 0.16$\pm$0.04    & $\le$0.04 & $\ge$10    & $---$   & \\
W of A&5 & 1.09$\pm$0.07    & 0.63$\pm$0.06    & 1.46$\pm$0.34    & 0.29$\pm$0.05    & $---$     & $---$     & $---$     & \\

\tablenotetext{a}{These line ratios are uncorrected for extinction
and stellar absorption.  Note that
the 
dispersions in the line ratios for the three different observations
of the NGC 4410A nucleus and Knot $\#$A are larger than the statistical uncertainties,
probably because of slight positional offsets.
}
\tablenotetext{b}{Observation number. See Figure~\ref{fig1} and Table~\ref{Table2}.}
\tablenotetext{c}{An artifact on the CCD chip 
made measurement of the [O~II]3727
line impossible in some cases. In the blue spectrum from Observation $\#$4, Knot $\#$2 was spatially unresolved
from the nucleus, so we do not provide blue line ratios for this source.
}
\tablenotetext{d}{ The line ratios of [O~III]/[O~II] and H$\alpha$/H$\beta$ for observations 2-5 (but not 
Observation 1) would
be decreased by 6-10\% and 2-3\% respectively by the correction for differential refraction.}
\end{tabular}
\end{table}

\begin{table}
\caption{Estimates of Extinction$^a$ \label{Table4}}
\begin{tabular}{llcccccccccccc}\hline
ID/obs&&A$_{\rm V}$\\ 
\hline
N4410A&1&$-$0.16$\pm$0.3\\
&4&0.9$\pm$0.3\\
&5&1.3$\pm$0.1\\
N4410B&2 &$---$\\
Knot $\#$1&1&$-1.5\pm0.3$\\
&2 &$-$0.5$\pm$0.5\\
&3 &0.6$\pm$0.15\\ 
Knot $\#$2&4 &$---$\\
Knot $\#$3&3 &1.6$\pm$0.2\\
Knot $\#$4&2 &2.3$\pm$0.5\\
NW Arc&4 &$---$\\ 
SW of A&1 &2.5$\pm$1.0\\ 
E of A&5 &$\ge$3.5\\ 
W of A&5 &$---$\\ 
\tablenotetext{a}{A systematic uncertainty of 12\% in the 
H$\alpha$/H$\beta$ ratios was included in the error budget to
account for uncertainties in the Galactic extinction (~10\%) 
and the smaller uncertainty (2\%) 
in relative slit losses due to differential refraction. 
No correction for stellar absorption has been applied.}
\end{tabular}
\end{table}

\end{document}